\documentclass[final,5p,times,twocolumn]{elsarticle}
\usepackage{bm}
\usepackage{amsmath}
\usepackage{euscript}
\usepackage{graphics}
\usepackage{amssymb}
\usepackage{amssymb}
 \usepackage{amsthm}

\newcommand{\be}{\begin{equation}}
\newcommand{\e}{\end{equation}}
\newcommand{\chast}{\int_{-\infty}^{\infty}\frac{d\omega'}{2\pi}}
\newcommand{\chastA}{\int_{-\infty}^{\infty}\frac{d\omega''}{2\pi}}
\newcommand{\beml}{\begin{subequations}}
\newcommand{\eml}{\end{subequations}}
\newcommand{\beq}{\begin{eqnarray}}
\newcommand{\eq}{\end{eqnarray}}
\newcommand{\ba}{\begin{array}}
\newcommand{\bali}{\begin{align}}
\newcommand{\ali}{\end{align}}
\newcommand{\ea}{\end{array}}

\newcommand{\lt}{\left}
\newcommand{\rt}{\right}
\newcommand{\n}{\nonumber}
\newcommand{\vs}{\vec{\sigma}}
\newcommand{\s}{\sigma}
\newcommand{\la}{\langle}
\newcommand{\ra}{\rangle}

\newcommand{\im}{\,{\rm Im}\,}
\newcommand{\re}{\,{\rm Re}\,}

\newcommand{\Dp}{\Delta^{(+)}}
\newcommand{\Dm}{\Delta^{(-)}}

\newcommand{\ur}{\urcorner}
\newcommand{\llc}{\llcorner}

\begin{document}

\begin{frontmatter}



\title{Double scattering of intense laser light by two atoms}

\author[frei,minsk]{Vyacheslav Shatokhin}
\author[frei]{Tobias Geiger}
\author[frei]{Thomas Wellens}
\author[frei]{Andreas Buchleitner}

\address[frei]{Institute of Physics, Albert-Ludwigs University of Freiburg, Hermann-Herder-Str. 3, 79104 Freiburg, Germany}
\address[minsk]{Stepanov Institute of Physics, National Academy
of Sciences, Skaryna Ave. 68, 220072  Minsk, Belarus}

\begin{abstract}
This paper analyzes coherent backscattering of intense laser light by two randomly placed distant atoms.
Starting from the general two-atom master equation, we analytically derive the elastic and inelastic background and interference components of the
double scattering spectrum. By expressing the final results in terms of single-atom observables, the two-atom problem is shown to be equivalent to a description in terms of single atoms under bichromatic driving. 
\end{abstract}


\end{frontmatter}






\section{Introduction}
Coherent backscattering (CBS) of light is a phenomenon emerging in a
dilute disordered medium, whereupon multiply scattered
counter-propagating waves interfere constructively in a narrow
angular range around the backward direction, leading to enhanced
intensity of the scattered signal \cite{sheng}. CBS of light has
attracted ongoing interest
due to its close
connection to various aspects of interference-induced effects in
presence of disorder, such as weak localization \cite{albada85}, Anderson
localization \cite{storzer06} or random lasers \cite{wiersma08}.

Since 1999, CBS of light has been observed in cold atoms
\cite{labeyrie99}. Quantum scatterers are endowed with internal
structure that impacts the matter-light interaction already at the level of a
single atom and, consequently, influences multiple scattering.
On the one hand, in the elastic scattering regime
of low laser intensity, the degeneracy of
atomic dipole transitions can strongly reduce the CBS signal, and
needs to be taken into account into the theory in order to meet the
experiment \cite{jonckheere00,labeyrie03}. On the other hand, for
degenerate (e.g., in rubidium atoms) and non-degenerate (e.g. in strontium atoms \cite{bidel02})
transitions alike, increasing the intensity of the incident laser
field leads to atomic saturation accompanied by nonlinear
(multi-photon) inelastic scattering processes \cite{scully} reducing the interference contrast in CBS \cite{chaneliere04}.

While a thorough theory of CBS of light from degenerate atoms in the regime of elastic scattering already exists
\cite{mueller01,kupriyanov03}, a theory of CBS from saturated atoms is still in a rudimentary state, several years after the
``saturation'' experiments with strontium \cite{chaneliere04} and
rubidium \cite{balik05} atoms. The main obstacle for developing such a theory is that CBS becomes a multi-wave interference effect in the nonlinear scattering regime \cite{wellens08,wellens09b},
with more than two interfering amplitudes. A 
nonperturbative (with respect to the strength of the nonlinearity)
theory of nonlinear CBS that fully accounts for the multi-wave interference character, has so far been developed only for nonlinear classical scatterers  \cite{wellens08,wellens09b}, not including quantum effects resulting from multi-photon scattering processes in the regime of atomic saturation.

Among the present theoretical
approaches  to CBS from saturated atoms,
one  \cite{wellens04} is based on diagrammatic
scattering theory and, hence, viable to
treating the multiple scattering processes of arbitrary order. It
is, however, bound to a weakly inelastic regime, and, so far, has
been worked out for two incident laser photons \cite{wellens06}.
Other approaches are based on quantum optical master
\cite{shatokhin05,shatokhin06} or Langevin \cite{gremaud06}
equations, and treat in detail the response of individual atomic
scatterers to the incident field of arbitrary intensity, though are
restricted to a few atoms.

Recently, we have proposed a new approach to multiple scattering of light from distant atoms in the saturation regime \cite{wellens09}. This method, to which we shall refer to as 'the pump-probe'  approach,
unifies the potential of diagrammatic scattering theory
with that of quantum optical methods, and, as we believe, will enable
us to develop a multiple scattering theory of intense laser light
from
dilute atomic gases, where the distances between atoms are much larger than the laser wavelength.

Within the pump-probe approach, which has so far been elaborated in detail for two scalar atoms \cite{wellens09,geiger09}, the
configuration-averaged double scattering signal is extracted from single-atom observables (such as the spectral correlation function of the atomic Bloch vector), where each atom is subject to a bichromatic driving field, consisting of the laser field (pump) and the weak field scattered from the other atom (probe) whose frequency may differ from the laser frequency. The main assumption of the pump-probe approach is that both these fields can be modelled as classical fields. Intuitively, this is expected to be possible in the case of large distance between both atoms, where only a single photon is scattered from one atom to the other, and thus correlations between different photons emitted by the same atom play no role.

As the numerical data shows \cite{wellens09,geiger09}, the
double scattering
spectra calculated within this new method are in excellent
agreement with the accurate results following from the two-atom master equation expanded up to lowest nonvanishing order in the
inverse distance between both atoms. This shows that the ansatz of a classical probe field is indeed justified for large distances.
Furthermore, for the elastic component of the backscattered field, we have analytically proven the strict equivalence of these two apparently very
different approaches \cite{geiger09}. However, the full analytical proof including also the inelastic spectra has been missing.

In the present contribution, we fill in this gap. Starting from the
master equation for two scalar atoms, we derive the double
scattering CBS elastic and inelastic signals, and express them in terms of purely single-atom
quantities. Next, we establish
the equivalence with the result of the pump-probe approach, based on single-atom equations
under the bichromatic driving \cite{wellens09,geiger09}.

The structure of the paper is as follows. In the next section, we
introduce the master equation approach and define all quantities
of physical interest. In Secs.~\ref{sec:elastic} and \ref{sec:inelastic}
we deduce, respectively, the elastic and inelastic components of
the backscattered light from two atoms. In Sec.~\ref{sec:bichromatic} we interpret the obtained results by considering a physical setting including single atoms subjected to bichromatic driving. Our work is
concluded in Sec.~\ref{sec:conclusion}.

\section{Master equation approach}
\subsection{Formalism}
\label{sec:formalism}
Let us consider a toy model of CBS
consisting of
two scalar
two-level
atoms embedded in a common electromagnetic bath and driven by a quasi-resonant laser wave with a wave vector $\vec{k}_L$.
In the following, we will neglect the atomic
center-of-mass
motion and focus on the dynamics of the internal degrees of freedom. The standard method to find the atomic dynamics is a master equation approach \cite{lehmberg70,agarwal74}, whereupon the bath degrees of freedom are traced out leading to the equation of motion for the quantum mechanical expectation values of atomic variables.

It should be noted that a more general model taking into account the vectorial nature of the scattered field as well as
of the atomic transitions has already been studied in some detail
\cite{shatokhin05,shatokhin06,shatokhin07}. The reason why we are
addressing here the simpler scalar case is that we want to
analytically derive expressions for the power spectrum of light
coherently backscattered from two atoms, and compare them with the
results of \cite{wellens09} obtained for {\it scalar} atoms.
However, we stress that the tools developed in this work can be
generalized for arbitrary atomic transitions.

In the Heisenberg picture and in the frame rotating at the laser frequency, one obtains the following master equation
for the expectation value of an arbitrary atomic operator $Q$ \cite{lehmberg70}:
 \begin{align}
\langle\dot{Q}\rangle&=\sum_{j=1}^2\langle-i\delta[\sigma^+_j\sigma^-_j,Q]
-\frac{i}{2}[\Omega_j\sigma_j^++\Omega^*_j \sigma^-_j,Q]\label{meq} \\
&-\gamma(\sigma^+_j\sigma^-_jQ+Q\sigma^+_j\sigma^-_j-2\sigma^
+_jQ\sigma^-_j)\rangle
\nonumber\\
&+\sum_{j\neq
k=1}^2\left(T(x)\langle[\sigma_j^+Q,\sigma^-_k]
+~T^*(x)[\sigma^+_j, Q\sigma^-_k]\rangle\right).\n
\end{align}
Here, $\sigma^-_j=|1\rangle_j\langle 2|_j$
and $\sigma_j^+=|2\rangle_j\langle 1|_j$,
with $|1\rangle_j$ and $|2\rangle_j$ being respectively the ground and excited states of atom $j$, denote the atomic lowering and raising operators.
Furthermore,
$\Omega_{j}=\Omega e^{i\vec{k}_L\cdot\vec{
r}_j}$ is the Rabi frequency dependent on the atomic position $\vec{r}_j$, $\delta=\omega_L-\omega_0$ is the laser-atom detuning, and $\gamma$ is half the Einstein's
$A$ coefficient.  The lower line of Eq.~(\ref{meq}) describes the retarded dipole-dipole interaction dependent on the dimensionless parameter $x=\omega_0|\vec{r}_1-\vec{r}_2|/c$. Coherent backscattering can be observed in the dilute regime
$x\gg 1$, where the
atoms are located in the far field of each other and exchange a single photon.
Within our toy model, this corresponds to a perturbative treatment
of the dipole-dipole interaction up to second order in the interatomic coupling $T(x)$. The explicit form of the latter in the limit $x\gg 1$ can be presented as
 \be
T(x)=i\gamma\frac{3}{2}\frac{e^{-ix}}{x}.
\label{dipdip} \e

By inserting into Eq.~(\ref{meq}) operators from the complete
two-atom basis set of operators: $Q\in \{\vec{q}_1\otimes\vec{q}_2\}$, with \beq \vec{q}_j&=&(\hat{I}_j,\sigma^-_j,\sigma^+_j,\sigma^z_j)^T,\n\\
\hat{I}_j&=&\sigma^+_j\sigma^-_j+\sigma^-_j\sigma^+_j, \quad
\sigma^z_j=\sigma^+_j\sigma^-_j-\sigma^-_j\sigma^+_j,\eq
one obtains a closed linear system of
16
coupled equations for the atomic averages.
Since $\langle \hat{I}_1\otimes \hat{I}_2\rangle=1$ for every atomic state (due to normalization), this can be reduced to a 15 dimensional system for the two-atom Bloch vector
\be \la\vec{Q}\ra=(\la\vec{\sigma}_2\ra,\la\vec{\sigma}_1\ra,\la\vec{\sigma}_1\otimes\vec{\sigma}_2\ra)^T
\label{eq:vecQ},\e
with
\be
\vec{\s}_j=(\s^-_j,\s^+_j,\s^z_j)^T.\e
This system of equations has the following matrix representation:
\be\langle\dot{\vec{Q}}\rangle=(A+V)\langle \vec{Q}\rangle+\vec{L}_{+,0},\label{matr_eq1}\e where the matrices $A$ and $V$
 describe the evolution of independent and dipole-dipole interacting atoms, respectively:
$A$ (as well as the free vector $\vec{L}_{+,0}$) is generated by the two upper lines, and $V$ by the lower line of equation (\ref{meq}).
The explicit forms of the matrices $A$ and $V$ as well as vector $\vec{L}_{+,0}$ will be given in Sec.~\ref{smatrix}.

\subsection{Intensity of coherently backscattered light}
\label{sintensity}
The solution of Eq.~(\ref{matr_eq1}) not only yields the atomic evolution, but also the field radiated by the atoms.
For example, the average stationary scattered light intensity in the direction of the unit vector $\vec{n}$,
\be
\langle I(\vec{n})\rangle=\lim_{t\to\infty}\la E^{(-)}(t)E^{(+)}(t)\ra_{\vec{n}},\e
after omitting unimportant prefactors and neglecting retardation, is given by
\be
\langle I(\vec{n})\rangle=\sum_j\la\s^+_j\s^-_j\ra_{\rm SS}+\sum_{j,k\neq j}\la\s_j^+\s^-_k\ra_{\rm SS} e^{ik\vec{n}\cdot(\vec{r}_j-\vec{r}_k)}
\label{eq:in2},
\e
where the subscript means that the expectation value is evaluated in the steady state. Bearing in mind that,
in the following, all
quantum mechanical expectation values are to be evaluated in the steady state, we will henceforth lighten notation by dropping this subscript. 

We note that the light intensity, Eq.~(\ref{eq:in2}), radiated by two distant atoms has already been considered previously, for example in \cite{rist08} (where the two atoms are coupled by an additional optical element). However, in the following, our focus will not be on the total intensity, as given by Eq.~(\ref{eq:in2}), but, with view at the multiple scattering scenario from a dilute cloud of atoms, we are rather interested in the double scattering coherent backscattering intensity, which is extracted from Eq.~(\ref{eq:in2}) by expansion in second order in the atom-atom interaction and subsequent average over the positions of atoms.
Zeroth-, first-, and second-order terms of the perturbative expansion of Eq.~(\ref{matr_eq1}) in the coupling
$V$
will carry upper indices `(0)', `(1)', and `(2)', respectively.
That is, all quantities that will be studied here are quadratic in the dipole-dipole coupling, and will be supplied with superscript `(2)'. For example, the double scattering intensity is obtained from
\be
\langle I(\vec{n})\rangle^{(2)}=\sum_j\la\s^+_j\s^-_j\ra^{(2)}+\sum_{j,k\neq j}\la\s_j^+\s^-_k\ra^{(2)} e^{ik\vec{n}\cdot(\vec{r}_j-\vec{r}_k)}.\label{eq:d_Int}
\e
Eq.~(\ref{eq:d_Int}) does not yet correspond to the CBS intensity. The latter is obtained after the configuration averaging procedure, denoted as $\la \ldots\ra_{\rm conf.}$,
\begin{align}\la\langle I(\vec{n})\rangle^{(2)}\ra_{\rm conf.}&=\sum_j\la\la\s^+_j\s^-_j\ra^{(2)}\ra_{\rm conf.}\label{eq:CBS_Int}\\
&+\sum_{j,k\neq j}\la\la\s_j^+\s^-_k\ra^{(2)} e^{ik\vec{n}\cdot(\vec{r}_j-\vec{r}_k)}\ra_{\rm conf.},\n
\end{align}
and includes two terms: the background, or ladder intensity, given by the first term in the right hand side of Eq.~(\ref{eq:CBS_Int}), and the interference, or crossed intensity, given by the second part of Eq.~(\ref{eq:CBS_Int}). The background and interference CBS intensities in Eq.~(\ref{eq:CBS_Int}) are formed respectively by
co- and counter-propagating double scattering amplitudes, cf. Fig.~\ref{fig:ladd-cros} below.
The background intensity is radiated uniformly into all directions $\vec n$, whereas the interference part contributes only in the backscattering direction, with a small angular width $\Delta\theta\propto 1/x\ll 1$. Thus the full angle dependence of the coherent backscattering intensity is determined by the background intensity and the height of the interference peak in exact backscattering direction, i.e.
$\vec{n}$ parallel to $-{\vec k}_L$, which we will assume from now on.

The configuration averaging procedure
includes two stages. At one stage,  averaging over interatomic distances is performed. As a result, the oscillating terms $\propto T^2(x), (T^*(x))^2$ will vanish, while terms $\propto |T(x)|^2$ which vary smoothly with $x$, will be preserved.  Another stage of configuration averaging includes averaging over random orientations of the radius-vector connecting the atoms (to kill oscillations of $\Omega_j, \Omega_j^*$ and most of their combinations). Note, however, that the phase of oscillating terms $\propto \Omega_j\Omega_k^*$ cancels itself with that of the phase factor $e^{ik\vec{n}\cdot(\vec{r}_j-\vec{r}_k)}$ in the backwards direction (second term of Eq.~(\ref{eq:d_Int})), since $k\approx k_L$. This is precisely the interference CBS contribution surviving the disorder averaging. It will be shown below that the averaging over both, distance and orientation, can be performed analytically within the master equation approach.

Once the average CBS intensity is known, one can assess the main measure of phase coherence called the enhancement factor \cite{jonckheere00},
defined as the interference divided by the background intensity, where, as mentioned above, the interference intensity is evaluated in exact backward direction, where it assumes its maximum value.
In the presence of inelastically scattered photons, a more refined measure of phase coherence is provided by the spectrum of backscattered light \cite{shatokhin07}. Indeed, the 
background and interference components of the
CBS spectrum indicate the interference character  and magnitude as a function of the frequency of the backscattered photon; the enhancement factor follows from them after integrating over the whole frequency distribution.

\subsection{Spectrum of coherently backscattered light}
\label{subsec:spectrum}
The master equation (\ref{meq}) is obviously Markovian, consequently, due to the quantum regression theorem \cite{scully}, the multi-time atomic correlation functions obey the same equation as Eq.~(\ref{matr_eq1}) but with modified initial conditions and free vector.
Among these multi-time correlation functions, we will consider the stationary first-order temporal coherence function of scattering atomic dipoles:
\be
\Gamma_1(\tau)=\sum_{j,k}\la\la\s^+_j\s^-_k(\tau)\ra^{(2)}e^{i\vec{k}\cdot\vec{r}_{jk}}\ra_{\rm conf.}\label{eq:corr},
\e
where $\vec{k}\equiv k_L\vec{n}$, and $\vec{r}_{jk}\equiv \vec{r}_j-\vec{r}_k$.
\subsubsection{Elastic spectrum}
The factorized correlation function of atomic dipoles
\begin{align}
\Gamma_{1;{\rm el}}(\tau)&=\sum_{j,k}\Bigl\la\left(\la\s^+_j\ra^{(0)}\la\s^-_k(\tau)\ra^{(2)}+\la\s^+_j\ra^{(1)}\la\s^-_k(\tau)\ra^{(1)}\right.\Bigr.\n\\
&\Bigl.\left.+\la\s^+_j\ra^{(2)}\la\s^-_k(\tau)\ra^{(0)}\right)e^{i\vec{k}\cdot\vec{r}_{jk}}\Bigr\ra_{\rm conf.},
\end{align}
gives the elastic component of the spectrum ($\propto \delta(\omega-\omega_L)$) through the Laplace transform \cite{scully}
\be
S_{\rm el}(\omega)=\frac{\re}{\pi}\int_0^\infty
d\tau e^{i\omega\tau}\Gamma_{1;{\rm el}}(\tau),\e
since the temporal evolution of the average dipole lowering operators in the steady state reduces to $e^{-i\omega_L\tau}$. Then
\be
S_{\rm el}(\omega)=I^{(2)}_{\rm el}\delta(\omega-\omega_L),\label{eq:Sel}\e
with
\begin{align}
I^{(2)}_{\rm el}&=\sum_{j,k=1}^2\Bigl\la\left(\la\s^+_j\ra^{(0)}\la\s^-_k\ra^{(2)}+\la\s^+_j\ra^{(1)}\la\s^-_k\ra^{(1)}\right.\Bigr.\n\\
&\Bigl.\left.+\la\s^+_j\ra^{(2)}\la\s^-_k\ra^{(0)}\right)e^{i\vec{k}\cdot\vec{r}_{jk}}\Bigr\ra_{\rm conf.}.\label{eq:Sp_el}\end{align}
The expression inside $\la\ldots\ra_{\rm conf.}$
explicitly depends on
the coordinates of atoms $j$ and $k$. However, after the configuration averaging, the result must be entirely symmetric with respect to indices interchange $j\leftrightarrow k$. Hence, instead of Eq.~(\ref{eq:Sp_el}), we can use shortened expressions. Let us write these expressions down separately for the background ($L_{\rm el}$) and interference ($C_{\rm el}$) contributions, assuming exactly the backward direction of observation ($\vec{k}=-\vec{k}_L$):
\begin{align}
L_{\rm el}&=2\la\la\sigma_2^+\ra^{(0)}\la\sigma_2^-\ra^{(2)}+\la\sigma_2^+
\ra^{(1)}\la\sigma_2^-\ra^{(1)}\n\\
&+\la\sigma_2^+\ra^{(2)}\la\sigma_2^-\ra^{(0)}\ra_{\rm conf.},\label{eq:el-ladd-1}\\
C_{\rm el}&=2\la(\la\sigma_1^-\ra^{(0)}\la\sigma_2^+\ra^{(2)}
+\la\sigma_1^-\ra^{(1)}\la\sigma_2^+\ra^{(1)}\n\\
&+\la\sigma_1^-\ra^{(2)}\la\sigma_2^+\ra^{(0)})e^{i\vec{k}_L\cdot\vec{r}_{12}}\ra_{\rm conf.}.\label{eq:el-cros-1}
\end{align}
Note that in Eq.~(\ref{eq:el-ladd-1}), we have fixed $j=k=2$, while in Eq.~(\ref{eq:el-cros-1}), $j=2$ and $k=1$.

\subsubsection{Inelastic spectrum}
\label{sec:inelastic-1}
In order to evaluate the inelastic spectrum of the backscattered light, let us introduce the time-dependent vector
\be
\Delta\vec{s}_j(\tau)\equiv \vec{s}_j(\tau)-\la\s^+_j\ra\la\vec{Q}(\tau)\ra,\e
where $\vec{s}_j(\tau) \equiv \la\s^+_j\vec{Q}(\tau)\ra$.
According to the quantum regression theorem \cite{scully},
the vector
$\Delta\vec{s}_j$ obeys the equation
\be
 \Delta \dot{\vec{s}}_j =( A+V)\Delta\vec{s}_j\label{eq:vecSj}\e
 with the initial condition
 \be
 \Delta\vec{s}_j(0)=\la\s^+_j\vec{Q}\ra-\langle\sigma_j^+\rangle\langle
\vec{Q}\rangle.
\label{eq:init_Dsj}\e
The inelastic spectrum follows from the fluctuating part of the
correlation function, Eq.~(\ref{eq:corr}):
\begin{align} \Delta
\Gamma_{1;{\rm inel}}(\tau)&=
\Gamma_{1}(\tau)-\Gamma_{1;{\rm el}}(\tau)\n\\ & =
\langle [\Delta\vec{s}^{\;(2)}_1(\tau)]_4+[\Delta\vec{s}^{\;(2)}_2(\tau)]_1\label{corr_in}\\
&+[\Delta\vec{s}^{\;(2)}_1(\tau)]_1e^{-i\vec{k}_L\cdot\vec{r}_{12}}+[\Delta\vec{s}^{\;(2)}_2(\tau)]_4e^{i\vec{k}_L\cdot\vec{r}_{12}}\rangle_{\rm conf.}\n\end{align}
via
Fourier transformation
\be
S_{\rm inel}(\nu)=\frac{\re}{\pi}\int^\infty_0 d\tau e^{i\nu\tau}
\Delta
\Gamma_{1;{\rm inel}}(\tau),
\label{eq:Sinel}\e
where $\nu=\omega-\omega_L$
denotes the detuning with respect to the laser frequency, and the fact that Eq.~(\ref{matr_eq1}) is written in the rotating frame is taken into account.
The indices $[\dots]_1$ and $[\dots]_4$ in Eq.~(\ref{corr_in}) refer to the  components $\sigma_2^-$ and $\sigma_1^-$ of the 15 dimensional vector $\vec{Q}$, see Eq.~(\ref{eq:vecQ}).

Employing the same symmetry argument as we did when writing down the expression for the elastic
contributions, let us write down
the corresponding
expressions for the inelastic background and interference spectra:
\begin{align}
L_{\rm inel}(\nu)&=\frac{2\re}{\pi}\int^\infty_0 d\tau e^{i\nu\tau}\la[\Delta\vec{s}^{\;(2)}_2(\tau)]_1\ra_{\rm conf.}\label{eq:Linel}\\
C_{\rm inel}(\nu)&=\frac{2\re}{\pi}\int^\infty_0 d\tau e^{i\nu\tau}\la[\Delta\vec{s}^{\;(2)}_2(\tau)]_4e^{i\vec{k}_L\cdot\vec{r}_{12}}\rangle_{\rm conf.}\label{eq:Cinel}\end{align}
Eqs.~(\ref{eq:Linel}) and (\ref{eq:Cinel}) can be cast into a more convenient form by using Laplace transforms
(which exist also for
purely imaginary Laplace transform variable $z=-i\nu$ because, as a consequence of the fluctuation-dissipation theorem
\cite{landau-V}, the fluctuating parts of the atomic correlation functions are exponentially decaying with time).
One obtains
\begin{align}
L_{\rm inel}(\nu)&=\frac{2\re}{\pi}\la[\Delta\vec{\tilde{s}}_2^{\;(2)}(-i\nu)]_1\ra_{\rm conf.},\label{L-inel_lap}\\
C_{\rm inel}(\nu)&=\frac{2\re}{\pi}\la[\Delta\vec{\tilde{s}}_2^{\;(2)}(-i\nu)]_4e^{i\vec{k}_L\cdot\vec{r}_{12}}\ra_{\rm conf.},\label{C-inel-lap}
\end{align}
where $\tilde{x}(z)=\int_0^\infty dt \exp(-zt)x(t)$ indicates the Laplace image of $x(t)$.

It follows from Eqs.~(\ref{eq:el-ladd-1}), (\ref{eq:el-cros-1}, (\ref{L-inel_lap}) and (\ref{C-inel-lap}) that finding the elastic and inelastic spectra of CBS requires
the solution of Eqs.~(\ref{matr_eq1}) and (\ref{eq:vecSj}), expanded up to second order in the dipole-dipole interaction matrix $V$,
with subsequent configuration averaging of the result. Both of these tasks can be accomplished by exploiting the tensor structure of matrices $A$ and $V$ to be discussed in the next section.

\subsection{Structure of the evolution matrices}
\label{smatrix}

Let us now study the structure of the matrices $A$ and $V$ introduced in Eq.~(\ref{matr_eq1}), referring to the evolution of non-interacting and interacting atoms, respectively.

\subsubsection{Free evolution: matrix A}
\label{smatrixa}

With the ordering of the 15-dimensional vector defined by Eq.~(\ref{eq:vecQ}), the
matrix $A$ receives the following block structure
\be
A=\lt(\begin{array}{cc} M_+& 0\\
L_\times&M_{\times}\end{array}\rt).\label{eq:defA}\e
The block matrix
$M_+$ has dimensions $6\times 6$; it describes the individual  evolution of
atoms 2 and 1 independently of each other:
\be M_+=M_2\oplus M_1, \e with the
matrix $M_j$ being the standard Bloch matrix 
for a single atom,
generated by the upper two lines of Eq.~(\ref{meq}):
\be
M_j=\lt(\ba{ccc}-\gamma+i\delta&0&-i\Omega_j/2\\
0&-\gamma-i\delta&i\Omega_j^*/2\\-i\Omega_j^*&i\Omega_j&-2\gamma\ea\rt).\label{eq:BlochM}\e
We remind that the optical Bloch equation
for a single atom
reads \cite{carmichael}
\be
\la\dot{\vec{\s_j}}\ra=M_j\la\vec{\s_j}\ra+\vec{L}, \quad \vec{L}=(0,0,-2\gamma)^T.\label{eq:Bloch_j}\e
The $9\times 9$ block $M_\times$ describes the evolution of two-atom correlation functions of uncoupled atoms and, hence, reads
\be M_{\times}=M_1\otimes I_2+I_1\otimes M_2,\e where $I_j$ is the $3\times 3$ unit matrix. In the following, we will drop the indices of the matrices unless the latter differ for atoms 1 and 2. Accordingly, the non-diagonal $9\times 6$ block matrix $L_\times$ can be written as
\be L_\times=(\vec{L}\otimes I\;\;I\otimes\vec{L}), \label{eq:L_times}\e
with $\vec{L}$ defined in Eq.~(\ref{eq:Bloch_j}). Finally, it is appropriate to specify here the vector $\vec{L}_{+,0}$ introduced in Eq.~(\ref{matr_eq1}):
\be
\vec{L}_{+,0}=(\underbrace{\vec{L}^{\,T},\vec{L}^{\,T}}_{\equiv\vec{L}^{\,T}_+},\underbrace{0,\ldots,0}_{9\;\text{zeroes}})^T.\label{eq:vecL_+}\e
To conclude, we have defined all ingredients of the composite system of two non-interacting atoms in terms of single-atom Bloch vectors and their tensor products. We will next do the same for the interacting atoms.

\subsubsection{Dipole-dipole interaction: matrix V}
\label{smatrixv}

\begin{figure}
\includegraphics[width=5cm]{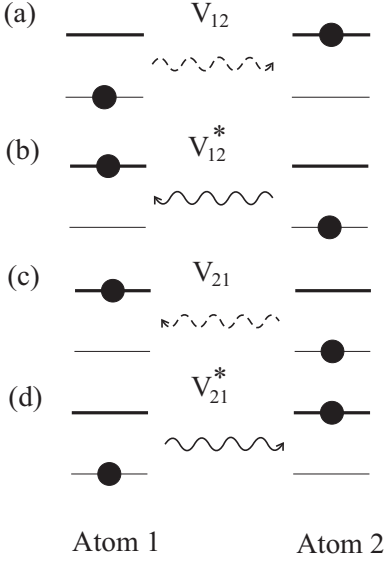}
\caption{Diagrammatic representation of the elementary dipole-dipole
interaction processes between two atoms (black spots correspond to their electronic states \emph{after} the interaction) with the respective terms of the matrix V, see
Eq.~(\ref{eq:matrV_1}). Solid (dashed) arrows depict positive
(negative) frequency photons. (a),(d): excitation is lost (gained)
by atom 1 (2); (b),(c): same in reversed order.}
\label{fig:processes}
\end{figure}

We remind that the matrix $V$ is generated by the third line of
Eq.~(\ref{meq}). It naturally splits into 4 components: \be
V=\underbrace{V_{12}+V_{21}}_{\propto
T(x)}+\underbrace{V_{12}^*+V_{21}^*}_{\propto
T^*(x)},\label{eq:matrV_1}\e where the first and second indices in the
subscripts coincide, respectively, with the values of the indices $j$ and
$k$ in Eq.~(\ref{meq}). The four matrices $V_{12}$, $V^*_{12}$,
$V_{21}$ and $V_{21}^*$ describe four elementary exchange excitation processes between the atoms which are depicted diagrammatically in Fig.~\ref{fig:processes}.
Furthermore, we associate terms $\propto
T^*(x)$ (or $T(x)$) with positive (or negative) frequency photons,
in
analogy with the single-atom case, see Eq.~(\ref{eq:Bloch_j}), where
$\la\s_j^-\ra$ (or $\la\s_j^+\ra$) are coupled to $\la\s_j^z\ra$ by
a coefficient proportional to the positive (or negative) frequency component of the laser field
$\Omega_j$ (or $\Omega_j^*$). Likewise,
in the
present two-atom case, it is easy to show that the expectation
values $\la\s_j^-\ra$ (or $\la\s_j^+\ra$) are coupled to $T^*(x)\la\s_j^-\s_k^z\ra$
(or $T(x)\la\s_j^+\s_k^z\ra$).

Let us proceed with the analysis of the structure of the matrix $V$. Each of the four components of the matrix $V$, Eq.~(\ref{eq:matrV_1}), has the following block structure:
 \be V_{jk}=
\lt(\begin{array}{cc} 0& (V_{jk})_\urcorner\\
(V_{jk})_\llcorner&(V_{jk})_\times\end{array}\rt)\label{eq:V_2},\e where
the dimensions of the blocks $(V_{jk})_\ur$, $(V_{jk})_\llc$, and $(V_{jk})_\times$ are
$6\times 9$, $9\times 6$, and $9\times 9$, respectively. From
Eqs.~(\ref{meq}), (\ref{eq:matrV_1}) and (\ref{eq:V_2}), it is
straightforward to
obtain explicit expressions for the matrices
$(V_{jk})_\alpha$ and $(V^*_{jk})_\alpha$ ($j\neq k=1,2$,
$\alpha=\ur,\llc,\times$). We will instead introduce them implicitly
-- through their action on probe vectors.

Let $\vec{a}_1$ and $\vec{a}_2$ be 3-component column vectors. Using these vectors we create two `probe' vectors
\be\vec{a}_1\otimes\vec{a}_2,\quad \lt(\ba{c}\vec{a}_2\\\vec{a}_1\ea\rt).\label{probe_vectors}\e
The interaction matrices are now characterized by the following identities:
\beml
\beq
(V_{12})_\ur(\vec{a}_1\otimes\vec{a}_2)&=&\lt(\ba{c}2iT\Dp\vec{a}_2(\vec{a}_1)_2\\\vec{0}\ea\rt),\label{id-V12-ur}\\
(V_{21})_\ur(\vec{a}_1\otimes\vec{a}_2)&=&\lt(\ba{c}\vec{0}\\2iT\Dp\vec{a}_1(\vec{a}_2)_2\ea\rt),\label{id-V21-ur}\\
(V_{12}^*)_\ur(\vec{a}_1\otimes\vec{a}_2)&=&\lt(\ba{c}\vec{0}\\-2iT^*\Dm\vec{a}_1(\vec{a}_2)_1\ea\rt),\label{id-V12*-ur}\\
(V_{21}^*)_\ur(\vec{a}_1\otimes\vec{a}_2)&=&\lt(\ba{c}-2iT^*\Dm \vec{a}_2(\vec{a}_1)_1\\
\vec{0}\ea\rt),\label{id-V21*-ur}\\
(V_{12})_\times(\vec{a}_1\otimes\vec{a}_2)&=&-2T\Dm\vec{a}_1\otimes\Dp\vec{a}_2,\label{id-V12-t}\\
(V_{21})_\times(\vec{a}_1\otimes\vec{a}_2)&=&-2T\Dp\vec{a}_1\otimes\Dm\vec{a}_2,\label{id-V21-t}\\
(V_{12}^*)_\times(\vec{a}_1\otimes\vec{a}_2)&=&-2T^*\Dm\vec{a}_1\otimes\Dp\vec{a}_2,\label{id-V12*-t}\\
(V_{21}^*)_\times(\vec{a}_1\otimes\vec{a}_2)&=&-2T^*\Dp\vec{a}_1\otimes\Dm\vec{a}_2,\label{id-V21*-t}\\
(V_{12})_\llc\lt(\ba{c}\vec{a}_2\\\vec{a}_1\ea\rt)&=&\vec{n}_1\otimes(2iT\Dp\vec{a}_2),\label{id-V12-ll}\\
(V_{21})_\llc\lt(\ba{c}\vec{a}_2\\\vec{a}_1\ea\rt)&=&2iT\Dp\vec{a}_1\otimes\vec{n}_1,\label{id-V21-ll}\\
(V_{12}^*)_\llc\lt(\ba{c}\vec{a}_2\\\vec{a}_1\ea\rt)&=& (-2iT^*\Dm\vec{a}_1)\otimes\vec{n}_2,\label{id-V12*-ll}\\
(V_{21}^*)_\llc\lt(\ba{c}\vec{a}_2\\\vec{a}_1\ea\rt)&=&\vec{n}_2\otimes(-2iT^*\Dm\vec{a}_2),\label{id-V21*-ll}
\eq
\label{eq:all_V}
\eml
where $(\vec{a}_1)_i$ and $(\vec{a}_2)_i$ (and similarly for other three-component vectors in the following) refer to the $i$-th component ($i=1,2,3$) in the basis given by the choice of the single-atom Bloch vector $(\la\s^-\ra,\la\s^+\ra,\la\s^z\ra)$. Furthermore,
the argument $x$ in the coupling constants is for brevity dropped, $\vec{n}_1=(\frac{1}{2},0,0)^T$, $\vec{n}_2=(0,\frac{1}{2},0)^T$, $\vec{0}=(0,0,0)^T$, and
\be
\Dm=\lt(\ba{ccc}0&0&-i/2\\0&0&0\\0&i&0\ea\rt),\;\;
\Dp=\lt(\ba{ccc}0&0&0\\0&0&i/2\\-i&0&0\ea\rt).\label{eq:defDmDp}
\e
Note that the matrices $V_\ur$, $V_\times$, and $V_\llc$ transform the probe vectors, Eq.~(\ref{probe_vectors}), to new vectors which can again be represented in the form of Eq.~(\ref{probe_vectors}).

As regards the matrices $\Dm$ and $\Dp$, they describe the coupling of the Bloch vector of an individual atom to another atom --
via coupling to
the two-atom correlation functions, and the coupling of the latter
back to the Bloch
vector of the other atom. Remarkably,
exactly the same matrices describe the coupling of a single atom to the classical laser field,
see the components proportional to $\Omega_j$ and $\Omega_j^*$ in Eq.~(\ref{eq:BlochM}).
This is not a mere coincidence. We will prove below that,
in the far-field limit, where the atoms exchange only single photon,
the radiation of one atom onto the other atom is equivalent to that of a classical field
with corresponding frequency distribution.

\section{The elastic component of CBS}
\label{sec:elastic}

As obvious from Eqs.~(\ref{eq:el-ladd-1}) and (\ref{eq:el-cros-1}), the elastic component of the spectrum results from the steady state of the two-atom Bloch vector, see Eq.~(\ref{matr_eq1}):
\be
\la\vec{Q}\ra=-(A+V)^{-1}\vec{L}_{+,0},\label{eq:StatSol}\e
expanded in different powers of the dipole-dipole interaction $V$ and finally configuration-averaged.

\subsection{Recurrence relations}
\label{srecurrence}

First, let us perform the expansion in powers of $V$. For this purpose, we split the 15-dimensional vector $\langle\vec{Q}\rangle$
into a 6- and a 9- dimensional vector characterizing, respectively, the single-atom Bloch vectors and the correlations between both atoms:
\be
\label{def:x-y}
\vec{x}^{\;(n)}\equiv \lt(\ba{c}\la\vec{\sigma}_2\ra\\
\la\vec{\sigma}_1\ra\ea\rt)^{(n)},\quad \vec{y}^{\;(n)}\equiv\la\vec{\sigma}_1\otimes\vec{\sigma}_2\ra^{(n)},
\e
where $n$ corresponds to the power in the series expansion in $V$.
By using the block structure of the matrix $A+V$ and applying to it the formula for the inversion of block matrices \cite{horn} (see \ref{sec:inversionA}), one can show that these vectors
 obey the following system of recurrence relations:
\beml
\label{recurrence}
\begin{align}
\vec{x}^{\;(n)}&=G_+V_\ur\vec{y}^{\;(n-1)}\label{eq:s2+s1-n}\\
\vec{y}^{\;(n)}&=G_\times V_\llc\vec{x}^{\;(n-1)}
+G_\times V_\times\vec{y}^{\;(n-1)}
+G_\times L_\times\vec{x}^{\;(n)}\label{eq:s1*s2-n},
\end{align}\eml
with $G_+=-M_+^{-1}$, $G_\times=-M_\times^{-1}$,
$V_\alpha=(V_{12})_\alpha+(V_{21})_\alpha+(V^*_{12})_\alpha+(V^*_{21})_\alpha$, $\alpha=\ur$, $\llc$, $\times$.
The initial condition reads:
\be
\vec{x}^{\;(0)}=\lt(\ba{c}G_2\vec{L}\\G_1\vec{L}\ea\rt),\;\; {\rm and}\;\; \vec{x}^{\;(n)}=0,\; \vec{y}^{\;(n)}=0,\;\;  {\rm for}\;\;n<0,
\label{x-0}
\e
where $G_j=-M_j^{-1}$. The relations (\ref{recurrence}) represent a specific case of the general recurrence relations for the sub-blocks of the matrix
$(z-A-V)^{-1}$, to be considered below in Sec.~\ref{sec:Inel-general}.

\subsection{Steady state solutions}
\label{sec:elastic1}

Using the above recurrence relations of Sec.~\ref{srecurrence}, we will now explicitly calculate the steady state for fixed atomic coordinates,
before we perform the configuration average in Sec.~\ref{sconfigav}.

From Eqs.~(\ref{def:x-y}) and (\ref{x-0}), we obtain the Bloch vectors for two non-interacting atoms ($n=0$):\be
\la\vs_2\ra^{(0)}=G_2\vec{L},\quad \la\vs_1\ra^{(0)}=G_1\vec{L}\label{eq:vs_2-0}.\e
The corresponding result for the correlations involves, according to Eq.~(\ref{eq:s1*s2-n}), the matrix $G_\times$, see also Eq.~(\ref{eq:s1*s2-0}).
For the sake of
reducing all calculations to the subspaces of atoms 1 and 2, we
shall use the following integral representation of the matrix
$G_\times$, proven in \ref{sec:int-repr-G}: \be
G_\times=\chast G_1(\pm
i\omega')\otimes
G_2(\mp i\omega')\label{eq:int_repr_G_times},\e where
$G_{1,2}(\pm i\omega')=(\pm i\omega'-M_{1,2})^{-1}$.
The  $\pm$ ($\mp$) signs of the integrands reflect invariance of the result, provided that the signs of the constituents are opposite.

Application of the integral representation Eq.~(\ref{eq:int_repr_G_times}) to Eq.~(\ref{eq:s1*s2-0}) yields:
\begin{align}
\la\vs_1\otimes\vs_2\ra^{(0)} & =\chast G_1(i\omega')\vec{L}\otimes G_2(-i\omega')G_2\vec{L}\n\\
& +\chast G_1(i\omega')G_1\vec{L}\otimes G_2(-
i\omega')\vec{L},\label{sum-rule-1} \end{align}
Eq.~(\ref{sum-rule-1}) can be evaluated by means of the general {\it sum rule} valid for $\re[z]=0$ (see proof in \ref{sec:sum-rule}):
\begin{align}
&\chast\ldots G_1\left(\frac{z}{2}+i\omega'\right)\ldots G_2\left(\frac{z}{2}-i\omega'\right)G_2(z)\ldots\n \\
+&\chast\ldots G_1\left(\frac{z}{2}+i\omega'\right)G_1\ldots G_2\left(\frac{z}{2}-i\omega'\right)\ldots\n\\
&=\ldots G_1\ldots G_2(z)\ldots,\label{eq:general_sum_rule}
\end{align}
where $\ldots$ stand for arbitrary expressions which do not depend on $\omega'$ and are
identical in all three lines of Eq.~(\ref{eq:general_sum_rule}). Applying this sum rule to Eq.~(\ref{sum-rule-1}) and using Eq.~(\ref{eq:vs_2-0}) yields:
\be\la\vs_1\otimes\vs_2\ra^{(0)}=
\la\vs_1\ra^{(0)}\otimes\la\vs_2\ra^{(0)},\label{vs1*vs2-0}\e
which is the obvious result for two non-interacting atoms.

The first-order
correction to the vector $\la\vs_2\ra^{(1)}$ can be found from
Eq.~(\ref{recurrence}).
Thereby,
 we arrive at \begin{align}
\la\vs_2\ra^{(1)}&=2iT_{12}(G_1\vec{L})_2G_2\Dp
G_2\vec{L}\n\\
&-2iT_{21}^*(G_1\vec{L})_1G_2\Dm G_2\vec{L},\label{eq:vs_2^1}\end{align}
where Eqs.~(\ref{id-V12-ur}), (\ref{id-V21*-ur}), and
(\ref{recurrence}) have been used. The first-order correction for the Bloch vector of the other atom results accordingly as:
\begin{align}\la\vs_1\ra^{(1)}&=2iT_{21}(G_2\vec{L})_2G_1\Dp
G_1\vec{L}\n\\
&-2iT_{12}^*(G_2\vec{L})_1G_1\Dm G_1\vec{L},\label{eq:vs_1^1}\end{align}
In order to be able to associate each term in Eqs.~(\ref{eq:vs_2^1}) and (\ref{eq:vs_1^1}) with the photon exchange processes depicted in Fig.~\ref{fig:processes}, we have supplemented the coupling strength $T$ with the corresponding indices, although their values are identical, i.e. $T_{12}=T_{21}=T$. As evident from Fig.~\ref{fig:processes}, and in agreement with Eqs.~(\ref{eq:vs_2^1}) and (\ref{eq:vs_1^1}), atom 2 is affected by processes (a) and (d), corresponding to $T_{12}$ and $T_{21}^*$,
whereas atom 1 is affected by (b) and (c), corresponding to $T_{21}$ and $T_{12}^*$,

Similarly, we have derived the first- and second-order corrections to the correlation functions $\la\vs_1\otimes\vs_2\ra^{(1)}$ and $\la\vs_2\ra^{(2)}$ (see \ref{sec:explicit-corrections}). Concerning the function $\la\vs_1\ra^{(2)}$, it follows from the expression (\ref{eq:vs_2^2}) for the function $\la\vs_2\ra^{(2)}$ after interchange of the lower indices $1\leftrightarrow 2$ numbering the atoms.

\subsection{Configuration average: ladder and crossed intensity}
\label{sconfigav}
Now, we have all ingredients to derive the elastic background and interference contributions $L_{\rm el}$ and $C_{\rm el}$ to the average backscattered intensity given by Eqs.~(\ref{eq:el-ladd-1}) and (\ref{eq:el-cros-1}). Indeed, these ingredients, $\la\vs_{1,2}\ra^{(0)}$, $\la\vs_{1,2}\ra^{(1)}$, and $\la\vs_{1,2}\ra^{(2)}$, are given by Eqs.~(\ref{eq:vs_2-0}), (\ref{eq:vs_2^1}), (\ref{eq:vs_1^1}) and (\ref{eq:vs_2^2}), respectively. We should finally perform configuration averaging of these expressions  over the coordinates of both atoms.

First, let us remind that both $L_{\rm el}$ and $C_{\rm el}$ are of second order in the interatomic couplings $T(x)$ and $T^*(x)$.
These couplings rapidly oscillate as a function of the distance
$r_{12}=x/k$ between the atoms, see Eq.~(\ref{dipdip}).
 Hence, as already mentioned in Sec.~\ref{sintensity}, only terms proportional to $|T(x)|^2$ survive averaging over the interatomic distance, whereas all terms proportional to
 $T(x)^2$ or $T^*(x)^2$ vanish.

 The next step is the average over the interatomic angle variables.
Also this averaging can be performed analytically, thanks to the following observation: the expression to be configuration averaged contains sums of products of matrix elements related to either of the atoms. It turns out (see \ref{sec:phases}) that all these matrix elements coincide, up to a position-dependent phase, with the ones obtained for an atom placed at the coordinate origin. This property reflects the fact that changing atomic positions changes nothing but the phases of the Rabi frequencies at the positions of atoms 1 and 2.
The disorder average is then survived only by those terms who have not gotten the phase factor; all the rest terms vanish.

Using the phase relations of \ref{sec:phases}, we can prove the following simple recipe in order to identify the terms surviving the configuration average: we remind that, according to Eq.~(\ref{eq:matrV_1}), $T$ and $T^*$ indicate the exchange of negative and positive frequency photons between the atoms. Moreover, the intensity of backscattered light, see Eq.~(\ref{eq:in2}), results from emission of a positive and a negative frequency photon from atom $1$ or $2$, indicated by the terms $\sigma_j^-$ and  $\sigma_j^+$, $j=1$ or $2$, respectively. As we have found, exactly those processes survive the ensemble average, where an atom emitting a positive (or negative) frequency photon has received this positive (or negative) frequency photon from the other atom. For this reason, we have labelled in Sec.~\ref{sec:elastic1} the coupling strengths
$T_{12}$, $T_{12}^*$, $T_{21}$ and $T_{21}^*$ in correspondence with the four processes depicted in Fig.~\ref{fig:processes}.
In particular, the ladder component, which results, according to Eq.~(\ref{eq:el-ladd-1}) from emission of a positive and a negative frequency photon from atom 2, involves only processes (a) and (d) of Fig.~\ref{fig:processes}. The disorder average is hence performed by extracting, among all terms
obtained when inserting Eqs.~(\ref{eq:vs_2-0}), (\ref{eq:vs_2^1}) and (\ref{eq:vs_2^2}) into Eq.~(\ref{eq:el-ladd-1}),
those which are proportional to $T_{12}T_{21}^*$.  The crossed contribution, involving emission of a positive frequency photon from atom 1 and a negative frequency photon from atom 2, consists of all the terms proportional to $T_{12}T_{12}^*$, corresponding to processes (a) and (b) in Fig.~\ref{fig:processes}.

Introducing notations:
\beml
\label{notations}
\begin{align}
\la\Delta\vs(\pm i\omega)\ra^{(\pm)}&\equiv
G(\pm i\omega)\Delta^{(\pm)} G\vec{L} \label{notation-a}\\
\la\Delta\vs(\omega)\ra^{(2)}&\equiv G\Dp G(-i\omega)\Dm G\vec{L}\n\\
&+G\Dm G(i\omega)\Dp G\vec{L}\label{notation-b},
\end{align}
\eml
and dropping the common prefactor $8T_{12}T_{21}^*$, we arrive at the following final expression for the elastic background intensity
\begin{align}
L_{\rm el}&=\la\s^+\ra^{(0)}\la\s^-\ra^{(0)}\bigl(\la\Delta\s^+(0)\ra^{(2)}\la\s^-\ra^{(0)}\bigr.\n\\
&\bigl.+\la\Delta\s^-(0)\ra^{(2)}\la\s^+\ra^{(0)}+\la\Delta\s^+(0)\ra^{(+)}\la\Delta\s^-(0)\ra^{(-)}\bigr.\n\\
&\bigl.+\la\Delta\s^+(0)\ra^{(-)}\la\Delta\s^-(0)\ra^{(+)}\bigr)\n\\
&+\chast P^{(0)}(\omega')\bigl(\la\Delta\s^+(\omega')\ra^{(2)}\la\s^-\ra^{(0)}\bigr.\n\\
&\bigl.+\la\Delta\s^-(\omega')\ra^{(2)}\la\s^+\ra^{(0)}\bigr),
\label{el-Ladder-final}
\end{align}
where \be
P^{(0)}(\omega)=(G(-i\omega)\vec{G}^{(0)}_1)_2+(G(i\omega)\vec{G}^{(0)}_2)_1,\label{Mollow}\e
with 
\beml
\label{vec-G10-G20}
\beq
\vec{G}_{1}^{(0)}&=&-i\Dp G\vec{L}+\vec{n}_2-(G\vec{L})_1G\vec{L},\label{eq:G1}\\
\vec{G}_{2}^{(0)}&=&+i\Dm G\vec{L}+\vec{n}_1-(G\vec{L})_2G\vec{L},\label{eq:G2}
\eq
\eml
gives the inelastic spectrum of single-atom resonance fluorescence
\cite{mollow69} known also as the Mollow triplet (see also \ref{sec:mollow}).

For the interference contribution, we obtain similarly:
\begin{align}
C_{\rm el}&=\la\s^+\ra^{(0)}\la\s^-\ra^{(0)}\la\Delta\s^-(0)\ra^{(-)}\la\Delta\s^+(0)\ra^{(+)}\n\\
&+\la\s^+\ra^{(0)}\la\s^+\ra^{(0)}\la\Delta\s^-(0)\ra^{(-)}\la\Delta\s^-(0)\ra^{(+)}\n\\
&+\la\s^-\ra^{(0)}\la\s^-\ra^{(0)}\la\Delta\s^+(0)\ra^{(-)}\la\Delta\s^+(0)\ra^{(+)}\n\\
&+\la\s^-\ra^{(0)}\chast\la\Delta\s^+(-i\omega')\ra^{(-)}(G\Dp
G(i\omega')\vec{G}^{(0)}_1)_2\n\\
&+\la\s^+\ra^{(0)}\chast(G\Dm G(-i\omega')\vec{G}^{(0)}_2)_1\la\Delta\s^-(i\omega')\ra^{(+)},
\label{el-Crossed-final}
\end{align}

\section{Inelastic spectrum of CBS}
\label{sec:inelastic}
\subsection{General remarks}
\label{sec:Inel-general}
In the previous section, we have presented a detailed derivation of the elastic spectrum of CBS backscattered from two laser-driven atoms. The analytical results, Eqs.~(\ref{el-Ladder-final}) and (\ref{el-Crossed-final}),
for the elastic ladder and crossed component
are manifestly single-atom expressions. While the physical setting leading to such expressions will be discussed below, here we will proceed with the inelastic spectrum. Although we will need essentially the same tools as we used for the derivation of the elastic spectrum, the pathway to the final results is
much lengthier for the inelastic case.
Henceforth, we will restrict ourselves to a sketch of the derivation.

In Sec.~\ref{sec:elastic}, it was shown that the background and
interference elastic intensities as expressed by
Eqs.~(\ref{eq:el-ladd-1},\ref{eq:el-cros-1}) correspond to processes
(a),(d) and (a),(b) of Fig.~\ref{fig:processes}, respectively.
The expressions given by Eqs.~(\ref{L-inel_lap}) and (\ref{C-inel-lap})
generalize Eqs.~(\ref{eq:el-ladd-1}) and (\ref{eq:el-cros-1}) to account
for inelastic scattering,
whereas the excitation exchange processes
remain the same. In other words, the ladder
and crossed  inelastic spectra, Eqs.~(\ref{L-inel_lap}) and  (\ref{C-inel-lap}), are also
bilinear forms in the matrices $V_{12}$, $V_{21}^*$ and $V_{12}$,
$V_{12}^*$, respectively.

Hence, to obtain the inelastic spectrum of CBS, we will
solve Eq.~(\ref{eq:vecSj}) for $j=2$ using Laplace transformation, with
the initial conditions being given by Eq.~(\ref{eq:init_Dsj}),
perturbatively to second order in the matrix $V$.

\subsubsection{Perturbative expansion of the resolvent}
The perturbative solution of Eq.~(\ref{eq:vecSj}) reads:
\begin{align}
\label{eq:R_n-s-m}
\Delta\vec{\tilde{s}}^{\;(2)}_2(z)&=R^{(0)}(z)\Delta\vec{s}^{\;(2)}_2(0)
+R^{(1)}(z)\Delta\vec{s}^{\;(1)}_2(0)\n\\&+
R^{(2)}(z)\Delta\vec{s}^{\;(0)}_2(0),\end{align} where $R^{(l)}(z)$,
$l=0,1,2$, represent subsequent terms in the series expansion of
the resolvent matrix \be R(z)=(z-A-V)^{-1}\e in $V$. Obviously, the only difference of the perturbative expansion of $R(z)$ from that of the matrix $(-A-V)^{-1}$ is that, in the latter case, the Green matrices are evaluated at $z=0$, while in the former case at $z=-i\nu$.
Accordingly, the four sub-blocks of the matrix $R(z)$: \be
R(z)=\lt(\ba{cc}z-M_+&-V_\ur\\
-L_\times-V_\llc&z-M_\times-V_\times\ea\rt)^{-1}\e satisfy the
recurrence relations generalizing Eqs.~(\ref{recurrence}):
\beml\label{recur_R}
\begin{align}
R^{(n)}_{11}(z)&=G_+(z) V_\ur R^{(n-1)}_{21}(z),\\
R^{(n)}_{12}(z)&=G_+(z) V_\llc R^{(n-1)}_{22}(z),\\
R^{(n)}_{2k}(z)&=G_\times L_\times R^{(n)}_{1k}(z)+G_\times V_\times
R^{(n-1)}_{2k}(z)\n\\
&+G_\times V_\llc R^{(n-1)}_{1k}(z),\quad k=1,2,\end{align}\eml with
the initial condition \be R^{(0)}_{11}(z)=G_+(z),\quad
R^{(0)}_{22}(z)=G_\times(z),\e and $R^{(n)}_{kl}=0$ for $n<0$.

\subsubsection{Initial conditions}
We will next show that the vectors of the initial conditions $\Delta\vec{s}^{(2-n)}_2(0)$ on which the sub-blocks of $R^{(n)}(z)$ act have the same structure as the probe vectors introduced in Eq.~(\ref{probe_vectors}).

Indeed, from the definition (\ref{eq:init_Dsj}) of $\Delta\vec{s}_2(0)$, one obtains \be
 \Delta\vec{s}^{\;(n)}_2(0)=\la\s^+_2\vec{Q}\ra^{(n)}-(\langle\sigma_2^+\rangle\langle
\vec{Q}\rangle)^{(n)}, \e what leads to the following expressions
\beml\label{init-expand}
\begin{align}
 \Delta\vec{s}^{\;(0)}_2(0)&=\la\s^+_2\vec{Q}\ra^{(0)}-\langle\sigma_2^+\rangle^{(0)}\langle
\vec{Q}\rangle^{(0)},\\
\Delta\vec{s}^{\;(1)}_2(0)&=\la\s^+_2\vec{Q}\ra^{(1)}-\langle\sigma_2^+\rangle^{(0)}\langle
\vec{Q}\rangle^{(1)}\n\\
&-\langle\sigma_2^+\rangle^{(1)}\langle
\vec{Q}\rangle^{(0)},\\
\Delta\vec{s}^{\;(2)}_2(0)&=\la\s^+_2\vec{Q}\ra^{(2)}-\langle\sigma_2^+\rangle^{(2)}\langle
\vec{Q}\rangle^{(0)}\n\\
&-\langle\sigma_2^+\rangle^{(0)}\langle\vec{Q}\rangle^{(2)}-\langle\sigma_2^+
\rangle^{(1)}\langle
\vec{Q}\rangle^{(1)}.
\end{align}
\eml
The vectors $\la\vec{Q}\ra^{(n)}$ are manifestly of the form (\ref{probe_vectors}) by virtue of the relations (\ref{eq:all_V}), (\ref{recurrence}), and (\ref{eq:general_sum_rule}). The same holds for the vector $\la\s_2^+\vec{Q}\ra^{(n)}$ which is defined as
\be
\la\s_2^+\vec{Q}\ra^{(n)}=\lt(\ba{c}i\Dm\la \vs_2\ra^{(n)}\\
\la\vs_1\s_2^+\ra^{(n)}\\
\la\vs_1\otimes i\Dm\vs_2\ra^{(n)}\ea\rt). \label{relation-s2Q}\e
Hence, the relations (\ref{eq:all_V}), (\ref{recurrence}), and (\ref{recur_R}), supplemented by the integral representation (\ref{eq:int_repr_G_times}) together with the identities derived in \ref{sec:sum-rule}, are sufficient to evaluate all terms in Eq.~(\ref{eq:R_n-s-m}).

\subsubsection{Configuration averaging}
The configuration averaging is reduced to the same procedure as for the elastic intensities. By choosing terms in the ladder and crossed spectra that are bilinear forms in the matrices $V_{12}$, $V_{21}^*$ and $V_{12}$,
$V_{12}^*$, respectively, and keeping only terms proportional to $|T|^2$, we automatically select the components surviving the disorder averaging. The averaged expressions for $\la[\Delta\vec{s}_2^{\;(2)}(-i\nu)]_1\ra_{\rm conf.}$ and $\la[\Delta\vec{s}_2^{\;(2)}(-i\nu)]_4e^{i\vec{k}_L\cdot\vec{r}_{12}}\ra_{\rm conf.}$  then follow from $[\Delta\vec{s}_2^{\;(2)}(-i\nu)]_1$  and $[\Delta\vec{s}_2^{\;(2)}(-i\nu)]_4$, respectively, after dropping the atomic indices.

\subsection{Inelastic ladder spectrum}
\label{sec:inel-ladder}
After configuration averaging of both sides of Eq.~(\ref{eq:R_n-s-m}),  we obtain
\begin{align}
\la[\Delta\vec{\tilde{s}}_2^{\;(2)}(-i\nu)]_1\ra_{\rm conf.}&=L^{(2;0)}(-i\nu)+L^{(1;1)}(-i\nu)\label{Ladder-inel-split}\\
&+L^{(0;2)}(-i\nu),\n
\end{align}
where
\be
L^{(2-n;n)}(-i\nu)=\la [R^{(2-n)}(-i\nu)\Delta\vec{s}^{\;(n)}_2(0)]_1\ra_{\rm conf.}.\e
In accordance with Eq.~(\ref{L-inel_lap}), the ladder spectrum reads
\be
L_{\rm inel}(\nu)=\frac{2\re}{\pi}(L^{(2;0)}(-i\nu)+L^{(1;1)}(-i\nu)+L^{(0;2)}(-i\nu)).\e
The explicit expressions for the functions $L^{(2;0)}(-i\nu)$, $L^{(1;1)}(-i\nu)$, and $L^{(0;2)}(-i\nu)$ are given in \ref{sec:Lnu-Cnu}.

Before we write down the final form of the inelastic ladder spectrum, let us introduce a new spectral function
\begin{align}
\label{P2_0}
P^{(2)}(\omega',\nu)&=2\re\Bigl\{(G(-i\nu)[\vec{G}_2^{(2)}(\omega')\Bigr.\n\\
&+\Dm G(-i\nu+i\omega')\Dp G(-i\nu)\vec{G}^{(0)}_{2}\n\\
&+\Dm G(-i\nu+i\omega')\vec{G}^{(+)}_{2}(i\omega')\n\\
&+\Dp G(-i\nu-i\omega')\Dm G(-i\nu)\vec{G}^{(0)}_{2}\n\\
&+\Bigl.\Dp G(-i\nu-i\omega')\vec{G}^{(-)}_{2}(-i\omega')])_1\Bigr\}.
\end{align}

By using the function $P^{(2)}(\omega',\nu)$ and the inelastic single-atom power spectrum $P^{(0)}(\omega')$, see Eq.~(\ref{Mollow}), we can write down the expression for the inelastic ladder spectrum in a compact form:
 \begin{align}
L_{\rm inel}(\nu)&=\frac{1}{2\pi}\chast P^{(0)}(\omega') P^{(2)}(\omega',\nu)\n\\
&+\frac{1}{2\pi}\la\s^+\ra^{(0)}\la\s^-\ra^{(0)}P^{(2)}(0,\nu)\n\\
&+\frac{1}{2\pi}\la\Delta\s^+(i\nu)\ra^{(-)}\la\Delta\s^-(-i\nu)\ra^{(+)}P^{(0)}(\nu)\n\\
&+\frac{1}{2\pi}\la\Delta\s^+(i\nu)\ra^{(+)}\la\Delta\s^-(-i\nu)\ra^{(-)}P^{(0)}(\nu).
\label{Lad-inel-final}
\end{align}
Not only has the expression for the inelastic background spectrum, Eq.~(\ref{Lad-inel-final}), become short thanks to the function $P^{(2)}(\omega',\nu)$. The latter function has a clear physical interpretation as
the cross section for a weak probe beam of frequency $\omega'$ to be scattered into the frequency $\nu$ by a strongly driven atom.
While this issue will be discussed in more detail
when introducing the pump-probe approach
in Sec.~\ref{sec:bichromatic} below, we will next derive the inelastic interference spectrum.

\subsection{Inelastic crossed spectrum}
\label{sec:inel-crossed}
Repeating the same steps as taken previously in the derivation of the inelastic background spectrum, we obtain
\begin{align}
\la[\Delta\vec{s}_2^{\;(2)}(-i\nu)]_4e^{i\vec{k}_L\cdot\vec{r}_{12}}\ra_{\rm conf.}&=C^{(2;0)}(-i\nu)+C^{(1;1)}(-i\nu)\n\\
&+C^{(0;2)}(-i\nu),\label{Crossed-inel-split}
\end{align}
where
\be
C^{(2-n;n)}(-i\nu)=\la [R^{(2-n)}(-i\nu)\Delta\vec{s}^{\;(n)}_2(0)]_4e^{i\vec{k}_L\cdot\vec{r}_{12}}\ra_{\rm conf.},\e
from which the inelastic spectrum reads
\be
C_{\rm inel}(\nu)=\frac{2\re}{\pi}(C^{(2;0)}(-i\nu)+C^{(1;1)}(-i\nu)+C^{(0;2)}(-i\nu))\label{Cinel-2},\e
where the explicit expressions for the functions $C^{(2;0)}(-i\nu)$, 
$C^{(1;1)}(-i\nu)$, and $C^{(0;2)}(-i\nu)$ are presented in \ref{sec:Lnu-Cnu}.

The crossed spectrum can conveniently be represented with
the help of the frequency correlation functions
$P^{(+)}(\omega',\nu)$ and
 $P^{(-)}(\omega',\nu)=(P^{(+)}(\nu,\omega'))^*$:
\begin{align}
P^{(+)}(\omega',\nu)&=(G(i\nu)\Dp G(i\nu-i\omega')\vec{G}^{(0)}_{1})_2\n\\
&+(G(i\nu)\vec{G}^{(+)}_{1}(i\omega'))_2\n\\
&+(G(-i\nu+i\omega')\Dp G(-i\nu)\vec{G}^{(0)}_{2})_1\n\\
&+(G(-i\nu+i\omega')\vec{G}^{(+)}_{2}(i\omega'))_1,\label{pplus}\\
P^{(-)}(\omega',\nu)&=(G(-i\nu)\Dm G(-i\nu+i\omega')\vec{G}^{(0)}_{2})_1\n\\
&+(G(-i\nu)\vec{G}^{(-)}_{2}(-i\omega'))_1\n\\
&+(G(i\nu-i\omega')\Dm G(i\nu)\vec{G}^{(0)}_{1})_2\n\\
&+(G(i\nu-i\omega')\vec{G}^{(-)}_{1}(-i\omega'))_2,\label{pminus}
\end{align}
The physical significance of the functions $P^{(+)}(\omega',\nu)$ and $P^{(-)}(\omega',\nu)$ as spectral correlation functions of a two-level system subjected to a bichromatic driving will be clarified below within the same setting as the function $P^{(2)}(\omega',\nu)$.
To this end, they allow us to write down a concise expression for the crossed inelastic spectrum as:
\begin{align}
C_{\rm inel}(\nu)&=\frac{1}{2\pi}\chast P^{(+)}(\omega',\nu)P^{(-)}(\nu-\omega',\nu)\n\\
&+\frac{1}{2\pi}\la\s^+\ra^{(0)}\la\Delta\s^-(-i\nu)\ra^{(-)}P^{(+)}(0,\nu)\n\\
&+\frac{1}{2\pi}\la\s^-\ra^{(0)}\la\Delta\s^+(i\nu)\ra^{(+)}P^{(-)}(0,\nu).
\label{Cros-inel-final}
\end{align}

\section{Pump-probe approach}
\label{sec:bichromatic} In Sections \ref{sec:elastic} and
\ref{sec:inelastic}, we derived the
configuration-averaged
double scattering elastic and
inelastic spectra of the background and interference contributions
to CBS from two atoms driven by a laser field of arbitrary
intensity. These quantities, given by Eqs.~(\ref{el-Ladder-final}),
(\ref{el-Crossed-final}), (\ref{Lad-inel-final}), and
(\ref{Cros-inel-final}), are expressed in terms of single-atom
averages and spectral correlation functions of which some, e.g.
$P^{(+)}(\omega',\nu)$, $P^{(-)}(\omega',\nu)$, and
$P^{(2)}(\omega',\nu)$ have been introduced rather formally. We
will next furnish them with a physical meaning. This meaning can be
straightforwardly clarified because precisely the same expressions
have been obtained by solving optical Bloch-like equations for a
single atom subject to bichromatic driving
\cite{wellens09,geiger09}.

Here, the driving field  ${\cal E}(t)$ is composed of
the injected laser field and the
photons
scattered by the other atom:
\begin{equation} {\cal
E}(t)={\cal E}_Le^{-i\omega_L t} +v^{(+)} e^{-i(\omega+\omega_L) t}
+{\cal E}_L^*e^{i\omega_L t}+ v^{(-)} e^{i(\omega+\omega_L) t}, \label{drive}
\end{equation}
both represented as a classical field with positive and negative frequency components. 
The laser field amplitude ${\cal E}_L$ is proportional to the Rabi frequency $\Omega$.
In correspondence to the approximation of large distance between the atoms, the probe fields $v^{(+)}$ and $v^{(-)}$ are assumed to be much weaker than the laser field ${\cal E}_L$. As already argued in the introduction, this approximation enables the modeling of the atomic radiation as a classical field acting on the other atom.
We then consider the frequency correlation function
\begin{align}
C(\omega_1,\omega_2) & = \int_{-\infty}^\infty \frac{dt_1dt_2}{2\pi}
  e^{-i(\omega_1+\omega_L) t_1+i(\omega_2+\omega_L) t_2}\nonumber\\
  & \times
 \left( \langle\sigma^+(t_1)\sigma^-(t_2)\rangle-\langle\sigma^+(t_1)\rangle\langle\sigma^-(t_2)\rangle\right),\label{dipole}
\end{align}
representing a positive frequency photon $\omega_1$ and a negative frequency photon $\omega_2$ (both with respect to the laser frequency) emitted by a single atom subject to
the bichromatic driving, Eq.~(\ref{drive}), and expand it in powers of the weak probe field amplitudes $v^{(+)}$ and $v^{(-)}$:
\begin{subequations}
\begin{align}
\left.C\right|_{v^{(\pm)}=0}
& =  \delta(\omega_1-\omega_2)P^{(0)}(\omega_1)\label{p0},\\
\left.\frac{\partial C}{\partial v^{(+)}}\right|_{v^{(\pm)}=0} & = 
\delta(\omega_2-\omega_1-\omega)
P^{(+)}(\omega,\omega_2),\ \ \ \ \ \label{c2} \\
\left.\frac{\partial C}{\partial v^{(-)}}\right|_{v^{(\pm)}=0}
& =  \delta(\omega_1-\omega_2-\omega)P^{(-)}(\omega,\omega_1)\label{c1},\\
\left.\frac{\partial^2 C}{\partial v^{(+)}\partial
    v^{(-)}}\right|_{v^{(\pm)}=0} &
=  \delta(\omega_1-\omega_2)P^{(2)}(\omega,\omega_1)\label{p1}.
\end{align}
\end{subequations}
The $\delta$-functions in Eqs.~(\ref{p0}-\ref{p1})
originate from integrating over $t_+:=t_1+t_2$
in Eq.~(\ref{dipole}), and are thus a consequence
of time translation invariance or energy conservation.
As we have found, the quantities defined by Eqs.~(\ref{p0}-\ref{p1}) -- i.e. by the perturbative solution of the single-atom Boch equations with bichromatic driving -- exactly correspond to the quantities $P^{(0)}$, $P^{(\pm)}$ and $P^{(2)}$ introduced above in Eqs.~(\ref{Mollow},\ref{P2_0},\ref{pplus},\ref{pminus}). This establishes the equivalence between the master equation and the pump-probe approach.

\begin{figure}
\includegraphics[width=8cm]{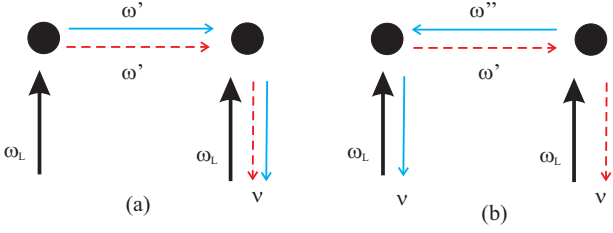}
\caption{Double scattering processes contributing to (a) ladder and
(b) crossed spectra of CBS. Thick arrows indicate (positive and
negative components of) the laser field driving the atoms (black spots),
thin solid (dashed) arrows indicate positive (negative) frequency
components of the fields scattered by the atoms. In all expressions,
the frequencies $\nu$, $\omega'$ and $\omega''$ are relative with
respect to the laser frequency $\omega_L$.} \label{fig:ladd-cros}
\end{figure}

Understanding the physical meaning of the quantities $P^{(0)}$, $P^{(\pm)}$ and $P^{(2)}$
becomes easier with the help of
diagrammatic language. Let us address a modified version of
Fig.~\ref{fig:processes} plotted in Fig.~ \ref{fig:ladd-cros}.
Namely, we added in Fig.~\ref{fig:ladd-cros} the laser field with
which both atoms continuously interact, the positive and negative
frequency components of the backscattered field, and combined the
exchange of excitation diagrams (a),(d) and (a), (b) of
Fig.~\ref{fig:processes} into the processes (a) and (b) of
Fig.~\ref{fig:ladd-cros}, respectively.
As discussed in Sec.~\ref{sconfigav}, these are the processes which survive the configuration average, after solving the two-atom master equation (\ref{meq}) perturbatively to second order in the interaction matrix $V$.

\begin{figure}
\includegraphics[width=6.5cm]{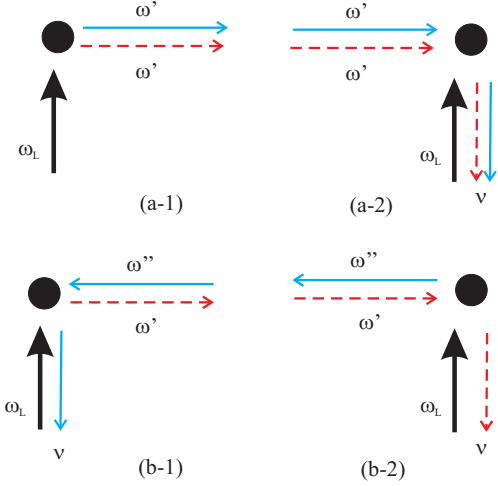}
\caption{Building blocks (diagrams) including the individual atoms with their
incoming and outgoing field. The combined diagrams (a-1) and (a-2) give
the ladder contribution (diagram (a) in Fig.~\ref{fig:ladd-cros}),
while the combined (b-1) and (b-2) diagrams give the crossed
contribution (diagram (b) in Fig.~\ref{fig:ladd-cros}).}\label{fig:splits}
\end{figure}

Now, let us split each of the diagrams (a) and (b) in
Fig.~\ref{fig:ladd-cros} into `building blocks', in order to
represent the individual atoms together with their incoming and outgoing
fields. The resulting 4 diagrams are shown in Fig.~\ref{fig:splits}.

Diagram (a-1) shows an atom subjected to a laser field.
Consequently, its outgoing fields correspond to resonance fluorescence with the
frequency distribution given by the function $P^{(0)}(\omega')$
defined in Eq.~(\ref{Mollow}). On diagram (a-2), an atom is subjected to
two classical fields: the laser field and a weak field at frequency
$\omega'$ scattered by the other atom. The response of the atom to such
a bichromatic field expanded to second order in the weak field
amplitude is described by the function $P^{(2)}(\omega',\nu)$
introduced in Eq.~(\ref{P2_0}). Integrating over the frequency $\omega'$
leads to $\int d\omega' P^{(0)}(\omega')P^{(2)}(\omega',\nu)$, which is the first contribution to the inelastic ladder intensity, Eq.~(\ref{Lad-inel-final}).
This is the contribution which originates when both atoms scatter inelastically, since the single-atom response as defined in
Eq.~(\ref{dipole}) above includes only inelastic scattering. However, the same analysis can be repeated including also elastic scattering, i.e. considering Eq.~(\ref{dipole}) without subtracting the
term $\langle\sigma^{(+)}\rangle\langle\sigma^{(-)}\rangle$ on the right hand side. As it turns out, the elastic response of a single atom to a probe field of frequency $\omega$ is then determined, in first and second order, by the corrections $\langle\Delta\vec{\sigma}(\pm i\omega)\rangle^{(\pm)}$ and $\langle\Delta\vec{\sigma}(\omega)\rangle^{(2)}$ to the Bloch vector defined in Eqs.~(\ref{notation-a},\ref{notation-b}). This yields the other terms of the inelastic ladder component, Eq.~(\ref{Lad-inel-final}), where one atom scatters inelastically and the other one elastically.

In the same spirit, the crossed spectrum is obtained as follows: diagram (b-1) describes a positive frequency component of the field
backscattered by the atom subjected to the laser field and to a positive
frequency component at the frequency $\omega''$ emitted by the other
atom. The atom also emits a negative frequency component at the
frequency $\omega'$ towards another atom. The frequency distribution of
these photons is described by the function $P^{(+)}(\nu,\omega')$. The frequency $\omega''$ is then determined, due to energy
conservation in the scattering processes, as $\omega''=\nu-\omega'$. Hence, diagram (b-2) is described by the
function $P^{(-)}(\nu-\omega',\nu)$. Integrating over $\omega'$ yields the first term of Eq.~(\ref{Cros-inel-final}), corresponding to inelastic-inelastic scattering, whereas the remaining terms are obtained by including a single elastic event.

\section{Conclusion}
\label{sec:conclusion} We have studied the spectrum of laser light doubly scattered between
two scalar atoms. By solving the two-atom master equation to second order
in the inverse distance between both atoms
and averaging over the random positions of the atoms, we derived
analytical expressions for the ladder (background) and crossed (interference) components of the coherent backscattering signal. The
results thereby obtained reduce to single-atom expressions coinciding exactly with
the results based on the pump-probe approach to coherent backscattering
\cite{wellens09,geiger09}. In the latter, heuristic, approach, all
correlation functions are calculated by considering from the very
beginning a single atom subject to a bichromatic
classical
driving field.

This does not contradict the well-known fact (see, for
instance, \cite{scully,carmichael}) that atomic radiation
exhibits photon antibunching and, hence, is nonclassical. Photon
antibunching and other nonclassical properties of resonance
fluorescence are quantum statistical properties which do not reveal
themselves inasmuch as single excitation exchange is involved.
Instead, these properties would become relevant for smaller interatomic distances, where several photons are exchanged between both atoms.
In the case of two distant atoms, corresponding to the experimentally realistic case of a dilute atomic medium, however, the classical modeling
is perfectly adequate
- at least in order to describe the intensity of the atomic radiation. Whether an appropriately extended pump-probe approach could also be able to reproduce 
 intensity-intensity correlations \cite{rist08,skornia01} is a subject for future studies.

As shown in this paper, the classical modeling of the fields exchanged between the atoms allows 
to reduce quantum mechanical calculations from the
two-atom Hilbert space to single-atom Hilbert spaces. In combination with multiple scattering theory as already existing for nonlinear classical scatterers \cite{wellens09,wellens09b}, the pump-probe approach thus opens the way to treat coherent backscattering of intense laser light by a
dilute cloud consisting of a very large number of atoms.

\section{Acknowledgements}

Funding by the DFG through an individual grant (V. S.) and through the Forschergruppe 760 is gratefully acknowledged.

\appendix
\section{Inversion of $A+V$}
\label{sec:inversionA}
Given a square matrix $B$, one can find its inverse by partitioning it into 4 blocks, the diagonal blocks being square matrices, and applying the formula [Ref. \cite{horn}, par. 0.7.3]
\begin{align}
B^{-1}&=\lt(\begin{array}{cc}b_1&b_2\\
b_3&b_4\end{array}\rt)^{-1}\label{eq:Schur_compl}\\
&=\lt(\ba{cc}
\lt[b_1-b_2b_4^{-1}b_3\rt]^{-1}&b_1^{-1}b_2\lt[b_3b_1^{-1}b_2-b_4\rt]^{-1}\\
\lt[b_3b_1^{-1}b_2-b_4\rt]^{-1}b_3b_1^{-1}&\lt[b_4-b_3b_1^{-1}b_2\rt]^{-1}\ea\rt).\n
\end{align}
For $B=A+V$,
\beml
\begin{align}
b_1&=M_+,\\
b_2&=V_\ur,\\
b_3&=L_\times+V_\llc,\\
b_4&=M_\times+V_\times.
\end{align}
\eml
Expanding blocks $B_{11}^{-1}$ and $B_{21}^{-1}$ to second order in $V$, we
establish the following relations
\beml \label{eq:recurr}\begin{align}
\lt(\ba{c}\la\vec{\sigma}_2\ra\\
\la\vec{\sigma}_1\ra\ea\rt)^{(0)}&=\lt(\ba{c}G_2\vec{L}\\G_1\vec{L}\ea\rt),\label{eq:s2+s1-0}\\
\la\vec{\sigma}_1\otimes\vec{\sigma}_2\ra^{(0)}&=G_\times L_\times\lt(\ba{c}\la\vec{\sigma}_2\ra\\
\la\vec{\sigma}_1\ra\ea\rt)^{(0)},\label{eq:s1*s2-0}\\
\lt(\ba{c}\la\vec{\sigma}_2\ra\\
\la\vec{\sigma}_1\ra\ea\rt)^{(1)}&=G_+V_\ur\la\vec{\sigma}_1\otimes\vec{\sigma}_2\ra^{(0)},\label{eq:s2+s1-1}\\
\la\vec{\sigma}_1\otimes\vec{\sigma}_2\ra^{(1)}&=G_\times V_\llc\lt(\ba{c}\la\vec{\sigma}_2\ra\\
\la\vec{\sigma}_1\ra\ea\rt)^{(0)}+G_\times V_\times\la\vec{\sigma}_1\otimes\vec{\sigma}_2\ra^{(0)}\n\\
&+G_\times L_\times\lt(\ba{c}\la\vec{\sigma}_2\ra\\\la\vec{\sigma}_1\ra\ea\rt)^{(1)},\label{eq:s1*s2-1}\\
\lt(\ba{c}\la\vec{\sigma}_2\ra\\\la\vec{\sigma}_1\ra\ea\rt)^{(2)}&=G_+V_\ur\la\vec{\sigma}_1
\otimes\vec{\sigma}_2\ra^{(1)},\label{eq:s2+s1-2}\\
\la\vec{\sigma}_1\otimes\vec{\sigma}_2\ra^{(2)}&=G_\times V_\llc\lt(\ba{c}\la\vec{\sigma}_2\ra\\\la\vec{\sigma}_1\ra\ea\rt)^{(1)}
+G_\times V_\times\la\vec{\sigma}_1
\otimes\vec{\sigma}_2\ra^{(1)}\n\\
&+G_\times L_\times\lt(\ba{c}\la\vec{\sigma}_2\ra\\
\la\vec{\sigma}_1\ra\ea\rt)^{(2)}\label{eq:s1*s2-2},\end{align}\eml where
$G_+=-M_+^{-1}=G_2\oplus G_1$, $G_\times=-M_\times^{-1}$,
with $G_j\equiv -M_j^{-1}$.
The recurrence relations (\ref{eq:recurr})  can be generalized to the form (\ref{recurrence}) by the method of induction.

\section{Integral representation of the Green's matrix $G_\times(z)$}
\label{sec:int-repr-G}
Since $M_\times=M_1\otimes I+I\otimes M_2$, we can write
\be e^{M_\times t}=e^{M_1 t}\otimes e^{M_2 t}.\label{eq:Mtens}
\e Laplace transformation applied to the left and right hand side of
Eq.~(\ref{eq:Mtens}) gives \be\frac{1}{z-M_\times}=\int_0^{\infty}dt
e^{-zt}e^{M_1 t}\otimes e^{M_2 t}. \quad \re[z]\geq
0\label{eq:LTtensM}\e In order to transform the right-hand side of
Eq.~(\ref{eq:LTtensM}) to tensor products of Laplace transforms of
individual atoms, we insert
$\delta(t-t')=\frac{1}{2\pi}\int_{-\infty}^{\infty}d\omega'
e^{i\omega'(t-t')}=\frac{1}{2\pi}\int_{-\infty}^{\infty}d\omega'
e^{i\omega'(t'-t)}$ to get: \be
G_\times(z)=\frac{1}{2\pi}\int_{-\infty}^{\infty}d\omega'
G_1\left(\frac{z}{2}\pm i\omega'\right)\otimes G_2\left(\frac{z}{2}\mp
i\omega'\right),\label{eq:int_rep}\e with \be
G_\times(z)=\frac{1}{z-M_\times},\quad G_j\left(\frac{z}{2}\pm
i\omega'\right)=\frac{1}{\frac{z}{2}\pm i\omega'-M_j}.\e

 For $z=0$, we obtain \be G_\times(0)\equiv G_\times=\chast G_1(\pm
i\omega')\otimes G_2(\mp i\omega'),\e which coincides with
Eq.~(\ref{eq:int_repr_G_times}).

When there are two Green's matrices, the generalization of Eq.~(\ref{eq:int_rep}) will be:  \begin{align} \ldots
G_\times(z)&\ldots G_\times(z)\ldots\equiv \chast\chastA\\
&\ldots G_1\lt(\frac{z}{2}\pm
i\omega'\rt)\ldots G_1\lt(\frac{z}{2}\pm i\omega''\rt)\ldots\n\\
&
\otimes
\ldots G_2\lt(\frac{z}{2}\mp i\omega'\rt)\ldots G_2\lt(\frac{z}{2}\mp
i\omega''\rt)\ldots. \n\end{align}

\section{Sum rule (\ref{eq:general_sum_rule}) and related identities}
\label{sec:sum-rule}
The eigenvalues $\lambda_k$ of the Bloch matrix $M_j$ given by
Eq.~(\ref{eq:BlochM}) are defined from the characteristic polynomial
\cite{mollow69} \be
f(z)=(z+2\gamma)\lt((z+\gamma)^2+\delta^2\rt)+(z+\gamma)\Omega^2.
\label{eq:characteristic}\e We note that $\lambda_k$ are independent
of the atomic position, and $\re[\lambda_k]<0$.

Solving the eigenvalue equation, one finds the right ($|u^j_k\ra$)
and left ($\la v_k^j|$) eigenvectors of the matrix $M_j$
(which
coincide with the right eigenvectors of the matrix $M_j^T$). Since
$\la v^j_k|u^j_q\ra=0$ for $k\neq q$ (see \cite{horn}, par.1.4.6), one can
represent $M_j$ as \be M_j=\sum_k \lambda_k P^j_k, \e where \be
P^j_k=\frac{|u^j_k\ra\la v^j_k|}{\la v^j_k|u^j_k\ra}, \quad \sum_k
P_k^j=I \e is the projector on the subspace corresponding to
the eigenvalue $\lambda_k$. One can easily deduce: \be
(z-M_j)=\sum_k (z-\lambda_k) P^j_k,\quad
G_j(z)=\sum_k\frac{P^j_k}{z-\lambda_k}\label{Green_projector}\e
With these preliminaries, we write down Eq.~(\ref{eq:general_sum_rule})
using Eq.~(\ref{Green_projector})
\begin{align}
(\ref{eq:general_sum_rule})&=\int_{-\infty}^{\infty}\frac{d\omega'}{2\pi}\ldots \sum_k\frac{P^1_k}{\omega'-i\frac{z}{2}+i\lambda_k}\n\\
&\times\ldots \sum_q\frac{P^2_q}{(\omega'+i\frac{z}{2}-i\lambda_q)(z-\lambda_q)}\ldots \n\\
&+\int_{-\infty}^{\infty}\frac{d\omega'}{2\pi}\cdot \sum_k\frac{P^1_k}{(\omega'-i\frac{z}{2}+i\lambda_k)(-\lambda_k)}\n\\
&\times \ldots \sum_q\frac{P^2_q}{\omega'+i\frac{z}{2}-i\lambda_q}\ldots. \label{eq:contourint}
\end{align}
These integrals can be easily calculated using the residues theorem.
The integrands in Eq.~(\ref{eq:contourint}) as a function of the complex variable $\omega'$ have two simple poles at $\omega_1'=iz/2-i\lambda_k$ and $\omega_2'=-iz/2+i\lambda_q$. Taking into account that $\re[z]=0$ and that $\re[\lambda_k]<0$ we note that $\im[\omega_1']>0$ and
$\im[\omega_2']<0$. By closing the integration contour in the upper half plane, we arrive after simple transformations at
\beq
(\ref{eq:contourint})&=&\ldots \sum_k\frac{P_k^1}{-\lambda_k}\ldots\ldots \sum_q\frac{P_q^2}{z-\lambda_q}\ldots\n\\
&=&\ldots G_1\ldots\ldots G_2(z)\ldots,\label{eq:proved-rule-1}\eq
which completes the proof of Eq.~(\ref{eq:general_sum_rule}).

As a concomitant result that we have used in our derivations,
we will present here the following identity:
\begin{align}
&\chast\chastA\ldots G_1\lt(-i\frac{\omega}{2}+i\omega''\rt)G_1(i\omega')\n\\&\ldots G_2\lt(-i\frac{\omega}{2}-i\omega''\rt)\ldots G_2(-i\omega')\ldots\label{sum-rule-II}\\
&=\chast\ldots G_1(i\omega')\ldots G_2(-i\omega-i\omega')\ldots G_2(-i\omega')\ldots\n\end{align}
The proof of Eq.~(\ref{sum-rule-II}) can be easily obtained by using the spectral decomposition, Eq.~(\ref{Green_projector}), and
direct evaluation of the integral over $\omega''$ with the help of the residues theorem.

In \ref{sec:Lnu-Cnu}, there appears a number of integrals which can be calculated by using a slight modification of the sum rule (\ref{eq:general_sum_rule}):
\begin{align}
\chast&\ldots G(z_1)G(i\omega')\ldots G(-i\omega')\ldots\n\\
+\chast&\ldots G(i\omega')\ldots G(z_2)G(-i\omega')\ldots\n\\
=&\ldots G(z_1)\ldots G(z_2)\ldots,\label{sum-rule-III}
\end{align}
where $z_1$ and $z_2$ are independent from $\omega'$ complex numbers
with $\re[z_1]=\re[z_2]=0$.

\section{Expressions for the functions $\la\vs_1\otimes\vs_2\ra^{(1)}$ and $\la\vs_2\ra^{(2)}$}
\label{sec:explicit-corrections}
Using Eqs.~(\ref{recurrence},\ref{vs1*vs2-0},\ref{eq:vs_2^1},\ref{eq:vs_1^1}), we obtain
\begin{align}
\la\vs_1\otimes\vs_2\ra^{(1)}=&-2 i T_{21}^*(G_1\vec{L})_1 G_1\vec{L}\otimes G_2\Dm G_2\vec{L}\n\\
&+2 iT_{12}(G_1\vec{L})_2G_1\vec{L}\otimes G_2\Dp G_2\vec{L}\n\\
&+2iT_{21}(G_2\vec{L})_2G_1\Dp G_1\vec{L}\otimes G_2\vec{L}\n\\
&-2iT^*_{12}(G_2\vec{L})_1G_1\Dm G_1\vec{L}\otimes G_2\vec{L}\n\\
&-2iT_{21}^*\chast G_1(i\omega')\vec{G}^{(0)}_{1;1}\n\\
&\otimes G_2(-i\omega')\Dm G_2\vec{L}\n\\
&+2iT_{12} \chast G_1(-i\omega')\vec{G}^{(0)}_{1;2}\n\\
&\otimes G_2(i\omega')\Dp G_2\vec{L}\n\\
&+2iT_{21}\chast G_1(i\omega')\Dp G_1\vec{L}\n\\
&\otimes G_2(-i\omega')\vec{G}^{(0)}_{2;2}\n\\
&-2iT_{12}^*\chast G_1(-i\omega')\Dm G_1\vec{L}\n\\
&\otimes G_2(-i\omega')\vec{G}^{(0)}_{2;1},
\label{eq:vs1*vs2-L-1}
\end{align}
where we have introduced new vectors
\beml
\label{vec-G010-G020}
\beq
\vec{G}_{j;1}^{(0)}&=&-i\Dp G_j\vec{L}+\vec{n}_2-(G_j\vec{L})_1G_j\vec{L},\label{eq:Gj1}\\
\vec{G}_{j;2}^{(0)}&=&+i\Dm G_j\vec{L}+\vec{n}_1-(G_j\vec{L})_2G_j\vec{L}.\label{eq:Gj2}
\eq
\label{eq:Gj1Gj2}
\eml
The vectors $\vec{G}^{(0)}_{j;1}$ and $\vec{G}^{(0)}_{j;2}$ play an important role in our derivations: they represent the initial conditions for correlation functions used in the calculations of the inelastic spectrum in single-atom resonance fluorescence (see \ref{sec:mollow}).

From Eqs.~(\ref{eq:general_sum_rule}), (\ref{eq:vs1*vs2-L-1}),(\ref{vec-G010-G020}),
 and (\ref{eq:s2+s1-n}), the second-order correction $\la\vs_2\ra^{(2)}$ for
 atom 2 follows straightforwardly. Here, we write down detailed expressions only
 for the terms proportional to $|T|^2$, and do not specify those proportional
 to $T^2$ or $(T^*)^2$ which will vanish after performing the configuration average,
 see Sec.~\ref{sconfigav}:
\begin{align}
\la\vs_2\ra^{(2)}=&+4T_{12}T_{21}^*(G_1\vec{L})_2(G_1\vec{L})_1
G_2\Dm G_2\Dp G_2\vec{L}\label{eq:vs_2^2}\\
&+4T_{12}T_{21}^*(G_1\vec{L})_2(G_1\vec{L})_1G_2\Dp G_2\Dm G_2\vec{L}\n\\
&+4T_{12}T_{12}^*(G_2\vec{L})_1(G_1\Dm G_1\vec{L})_2G_2\Dp G_2\vec{L}\n\\
&+4T_{21}T_{21}^*(G_2\vec{L})_2(G_1\Dp G_1\vec{L})_1G_2\Dm G_2\vec{L}\n\\
&+4T_{12}T_{21}^*\chast\lt(G_1(-i\omega')\vec{G}^{(0)}_{1;2}\rt)_1\n\\ &\times G_2\Dm G_2(i\omega')\Dp G_2\vec{L}\n\\
&+4T_{12}T_{21}^*\chast\lt(G_1(i\omega')\vec{G}^{(0)}_{1;1}\rt)_2\n\\ &\times G_2\Dp G_2(-i\omega')\Dm G_2\vec{L}\n\\
&+4T_{12}T_{12}^*\chast\lt(G_1(i\omega')\Dm G_1\vec{L}\rt)_2\n\\
&\times G_2\Dp G_2(-i\omega')\vec{G}^{(0)}_{2;1}\n\\
&+4T_{21}T_{21}^*\chast\lt(G_1(i\omega')\Dp G_1\vec{L}\rt)_1\n\\
&\times G_2\Dm
G_2(-i\omega')\vec{G}^{(0)}_{2;2} +(T^2)\dots+(T^*)^2\dots.\n
\end{align}
By reasons of symmetry, the corresponding result $\la\vs_1\ra^{(2)}$ for atom 1 is obtained by exchanging the Green's functions $G_1\leftrightarrow G_2$, the vectors $\vec{G}_{1;1}^{(0)}\leftrightarrow \vec{G}_{2;1}^{(0)}$ and $\vec{G}_{1;2}^{(0)}\leftrightarrow \vec{G}_{2;2}^{(0)}$,
 and the coupling strengths $T_{12}\leftrightarrow T
 _{21}$ and $T^*_{12}\leftrightarrow T^*_{21}$.
\section{Phase relations}
\label{sec:phases}
The following relations between expressions for an atom placed at position $\vec{r}_j$ (left-hand side) and
an atom placed at the origin (right-hand side) can be obtained by direct calculation:
\beml
\label{phases}
\begin{align}
(G_j\vec{L})_1&=(G\vec{L})_1e^{i\vec{k}_L\cdot\vec{r}_j},\\
(G_j\vec{L})_2&=(G\vec{L})_2e^{-i\vec{k}_L\cdot\vec{r}_j}\\
(G_j(z)\Dp G_j\vec{L})_1&=(G(z)\Dp G\vec{L})_1e^{2i\vec{k}_L\cdot\vec{r}_j},\\
(G_j(z)\Dp G_j\vec{L})_2&=(G(z)\Dp G\vec{L})_2\\
(G_j(z)\Dm G_j\vec{L})_1&=(G(z)\Dm G\vec{L})_1,\\
(G_j(z)\Dm G_j\vec{L})_2&=(G(z)\Dm G\vec{L})_2e^{-2i\vec{k}_L\cdot\vec{r}_j}\\
(G_j\Dm G_j(z)\Dp G_j\vec{L})_1&=(G\Dm G(z)\Dp G\vec{L})_1e^{i\vec{k}_L\cdot\vec{r}_j}\\
(G_j\Dm G_j(z)\Dp G_j\vec{L})_2&=(G\Dm G(z)\Dp G\vec{L})_2 e^{-i\vec{k}_L\cdot\vec{r}_j}\\
(G_j\Dp G_j(z)\Dm G_j\vec{L})_1&=(G\Dp G(z)\Dm G\vec{L})_1
e^{i\vec{k}_L\cdot\vec{r}_j}\\
(G_j\Dp G_j(z)\Dm G_j\vec{L})_2&=(G\Dp G(z)\Dm G\vec{L})_2 e^{-i\vec{k}_L\cdot\vec{r}_j}, \end{align} \eml where $G(z)\equiv
G_j(z)|_{\vec{r}_j=\vec{0}}$, and $z$ is an imaginary number.

Each of the terms to be averaged represents a product of matrix elements of the form given by (\ref{phases}) (see, for instance, Eq.~(\ref{eq:vs1*vs2-L-1})). Then, using (\ref{phases}), one can easily deduce the total phase of each term. There are two types of terms surviving the disorder averaging: (i) terms without the total phase factor and (ii) terms with the total phase factor $e^{-i\vec{k}_L\cdot\vec{r}_{12}}$. The former contribute to the ladder spectrum, and the latter  to the crossed spectrum. This rule can also be generalized for the inelastic spectrum.

\section{Inelastic spectrum in single-atom resonance fluorescence}
\label{sec:mollow}
The expression for the
inelastic spectrum, Eq.~(\ref{eq:Sinel}), applied to the single-atom case can be written as
\begin{align}
P^{(0)}(\nu)&=\frac{1}{2\pi}\int_0^{\infty}d\tau \Bigl(e^{i\nu\tau}\lt(\Delta\vec{s}_{j;+}(\tau)\rt)_1\Bigr.\\
&\Bigl.+e^{-i\nu\tau}\lt(\Delta\vec{s}_{j;-}(\tau)\rt)_2\Bigl),
\end{align}
where $\Delta\vec{s}_{j;\pm}(\tau)$ obeys the equation of motion
\be
\Delta\dot{\vec{s}}_{j;\pm}=M_j\Delta\vec{s}_{j;\pm},\label{eq:dsj_t}\e
with initial conditions
\beml
\begin{align}
\Delta\vec{s}_{j;+}(0)&=\la\s_j^+\vs_j\ra-\la\s_j^+\ra\la\vs_j\ra,\\
\Delta\vec{s}_{j;-}(0)&=\la\vs_j\s_j^-\ra-\la\s_j^-\ra\la\vs_j\ra.
\end{align}
\label{eq:sj+sj-}
\eml
Comparing Eqs.~(\ref{eq:sj+sj-}) and (\ref{eq:Gj1Gj2}) one obtains
\begin{align}
 \Delta\vec{s}_{j;+}(0)&=\vec{G}_{j;2}^{(0)},\\
\Delta\vec{s}_{j;-}(0)&=\vec{G}_{j;1}^{(0)}.
\end{align}
Solving Eq.~(\ref{eq:dsj_t}) with the help of the Laplace transform, we obtain the following expression for the inelastic spectrum of resonance fluorescence:
\be
P^{(0)}(\nu)=\frac{1}{2\pi}\Bigl(\lt(G_j(i\nu)\vec{G}^{(0)}_{j;2}\rt)_1
+\lt(G_j(-i\nu)\vec{G}^{(0)}_{j;1}\rt)_2\Bigr).\label{eq:intro-S0}\e
The spectral distribution of resonance fluorescence is independent of the coordinate of the atom. Accordingly, the atomic index $j$ can be  dropped in the right hand side of Eq.~(\ref{eq:intro-S0}), since
\beml
\label{phase_spectrum}
\begin{align}
 \lt(G_j(i\nu)\vec{G}^{(0)}_{j;2}\rt)_1&=\lt(G(i\nu)\vec{G}^{(0)}_{2}\rt)_1,\\
 \lt(G_j(-i\nu)\vec{G}^{(0)}_{j;1}\rt)_2&=\lt(G(-i\nu)\vec{G}^{(0)}_{1}\rt)_2,\end{align}
\eml
where the expressions with the dropped atomic index on the right hand side of Eq.~(\ref{phase_spectrum}) correspond to an atom placed at the coordinate origin.

\section{Spectral correlation functions $L^{(2-n;n)}(-i\nu)$ and $C^{(2-n;n)}(-i\nu)$}
\label{sec:Lnu-Cnu}
Omitting the common prefactor $4|\bar{T}|^2$, we get the following results for the functions $L^{(2-n;n)}(-i\nu)$:
\begin{align}
L^{(2;0)}(-i\nu)&=(G\vec{L})_1(G\vec{L})_2\lt(G(-i\nu)\Dp G(-i\nu)\Dm G(-i\nu)\vec{G}^{(0)}_{2}\rt)_1\n\\
&+(G\vec{L})_1(G\vec{L})_2\lt(G(-i\nu)\Dm G_2(-i\nu)\Dp G(-i\nu)\vec{G}^{(0)}_{2}\rt)_1\n\\
&+\chast\lt(G(i\omega')\vec{G}^{(0)}_{1}\rt)_2\n\\
&\times\lt(G(-i\nu)\Dp G(-i\nu-i\omega')\Dm G(-i\nu)\vec{G}^{(0)}_{2}\rt)_1\n\\
&+\chast\lt(G(-i\omega')\vec{G}^{(0)}_{2}\rt)_1\n\\
&\times\lt(G(-i\nu)\Dm G(-i\nu+ i\omega')\Dp G(-i\nu)\vec{G}^{(0)}_{2}\rt)_1,\label{L2;0}\end{align}
\begin{align}
L^{(1;1)}(-i\nu)&=(G\vec{L})_1(G\vec{L})_2\lt(G(-i\nu)\Dp G(-i\nu)\vec{G}^{(-)}_{2}(0)\rt)_1\n\\
&+(G\vec{L})_1(G\vec{L})_2\lt(G(-i\nu)\Dm G(-i\nu)\vec{G}^{(+)}_{2}(0)\rt)_1\n\\
&+\chast\lt(G(i\omega')\vec{G}^{(0)}_{1}\rt)_2\n\\
&\times\lt(G(-i\nu)\Dp G(-i\nu-i\omega')\vec{G}^{(-)}_{2}(-i\omega')\rt)_1\n\\
&+\chast\lt(G(-i\omega')\vec{G}^{(0)}_{2}\rt)_1\n\\
&\times\lt(G(-i\nu)\Dm G(-i\nu+i\omega')\vec{G}^{(+)}_{2}(i\omega')\rt)_1\n\\
&+\chast\lt(G(-i\nu)G(i\omega')\vec{G}^{(0)}_{1}\rt)_2\n\\
&\times\lt(G(-i\omega')\Dm G\vec{L})_2(G(-i\nu)\Dp G\vec{L}\rt)_1\n\\
&+\chast\lt(G(-i\nu)G(-i\omega')\vec{G}^{(0)}_{2}\rt)_1\n\\
&\times(G(i\omega')\Dp G\vec{L})_2(G(-i\nu)\Dm G\vec{L})_1,\label{L1;1}\end{align}
\begin{align}
L^{(0;2)}(-i\nu)&=(G\vec{L})_1(G\vec{L})_2\lt(G(-i\nu)\vec{G}^{(2)}_{2}(0)\rt)_1\n\\
&+\chast\lt(G(-i\omega')\vec{G}^{(0)}_{2}\rt)_1\lt(G(-i\nu)\vec{G}^{(2)}_{2}(\omega')\rt)_1\n\\
&+\chast\lt(G(i\omega')\vec{G}^{(0)}_{1}\rt)_2\lt(G(-i\nu)\vec{G}^{(2)}_{2}(\omega')\rt)_1\n\\
&+\chast\lt(G(-i\omega')\vec{G}^{(0)}_{2}\rt)_1\n\\
&\times\lt(G(-i\nu)G(-i\omega')\Dm G\vec{L}\rt)_1\lt(G(i\omega')\Dp G\vec{L}\rt)_2\n\\
&+\chast\lt(G(-i\omega')\vec{G}^{(0)}_{2}\rt)_1\n\\
&\times\lt(G(-i\nu)G(i\omega')\Dp G\vec{L}\rt)_1\lt(G(-i\omega')\Dm G\vec{L}\rt)_2\n\\
&+\chast\lt(G_1(i\omega')\vec{G}^{(0)}_{1}\rt)_2\n\\
&\times\lt(G(-i\nu)G(i\omega')\Dp G\vec{L}\rt)_1\lt(G(-i\omega')\Dm G\vec{L}\rt)_2\n\\
&+\chast\lt(G(i\omega')\vec{G}^{(0)}_{1}\rt)_2\n\\
&\times\lt(G(-i\nu)G(-i\omega')\Dm G\vec{L}\rt)_1\lt(G(i\omega')\Dp G\vec{L}\rt)_2,\label{L0;2}
\end{align}
where the vectors $\vec{G}^{(-)}_{2}(-i\nu)$, $\vec{G}^{(+)}_{2}(i\nu)$, and $\vec{G}^{(2)}_{2}(\nu)$ at $\omega=0$ are related to the vectors
of initial conditions $\Delta\vec{s}_2^{(2)}(0)$. They are defined as
\begin{align}
\vec{G}^{\;(-)}_{2}(-i\nu)&=i\Dm G(-i\nu)\Dm G\vec{L}-G(-i\nu)\Dm G\vec{L}\lt(G\vec{L}\rt)_2\n\\
&-G\vec{L}\lt(G(-i\nu)\Dm G\vec{L}\rt)_2,\label{G-2}\end{align}
\begin{align}
\vec{G}^{\;(+)}_{2}(i\nu)&=i\Dm G(i\nu)\Dp G\vec{L}-\lt(G(i\nu)\Dp G\vec{L}\rt)_2G\vec{L}\n\\
&-\lt(G\vec{L}\rt)_2G(i\nu)\Dp G\vec{L},\label{G+2}\end{align}
\begin{align}
\vec{G}^{\;(2)}_{2}(\nu)&=i\Dm G\Dp G(-i \nu)\Dm G\vec{L}+i\Dm G\Dm G(i \nu)\Dp G\vec{L}\n\\
&-G\vec{L}\lt(G\Dp G(-i \nu)\Dm G\vec{L}\rt)_2-G\vec{L}\lt(G\Dm G(i \nu)\Dp G\vec{L}\rt)_2\n\\
&-\lt(G\vec{L}\rt)_2G\Dp G(-i \nu)\Dm G\vec{L}-\lt(G\vec{L}\rt)_2G\Dm G(i \nu)\Dp G\vec{L}\n\\
&-\lt(G(i\nu)\Dp G\vec{L}\rt)_2G(-i\nu)\Dm G\vec{L}\n\\
&-\lt(G(-i\nu)\Dm G\vec{L}\rt)_2G(i\nu)\Dp G\vec{L}.\label{G2-2}
\end{align}
To obtain the complex conjugated part of any of the expressions from Eqs.~(\ref{L2;0},\ref{L1;1},\ref{L0;2},\ref{G-2},\ref{G+2},\ref{G2-2}) needed for calculation of their real parts, one has to make the following interchanges in these equations:
(a) $i \leftrightarrow -i$; (b) all subscripts $1\leftrightarrow 2$; (c) all superscripts $(+)\leftrightarrow (-)$. The superscripts $(0)$ and $(2)$ should be left unchanged. These interchanges imply introducing  3 new vectors, $\vec{G}_1^{\;(+)}(i\nu)$, $\vec{G}_1^{\;(-)}(-i\nu)$, and $\vec{G}_1^{\;(2)}(\nu)$, whose explicit expressions follow immediately from Eq.~(\ref{G-2},\ref{G+2},\ref{G2-2}) by applying the substitutions (a)-(c) specified above.

Next, we note that the 2 lower lines in Eq.~(\ref{L1;1}) and 4 lower lines in Eq.~(\ref{L0;2}) form, with their complex conjugates, pairs of terms for which integration over $\omega'$ can be performed analytically with the help of Eqs.~(\ref{eq:general_sum_rule}, \ref{sum-rule-II},\ref{sum-rule-III}). After these transformations, the corresponding 12 integrals are reduced to 4 algebraic expressions.

For the correlation functions entering Eq.~(\ref{Cinel-2}) we obtain
\begin{align}
C^{(2;0)}(-i\nu)&=(G\vec{L})_2\lt(G(-i\nu)\Dm G\vec{L}\rt)_1\lt(G(-i\nu)\Dp G(-i\nu)\vec{G}^{(0)}_{2}\rt)_1\n\\
&+\chast\lt(G(-i\nu)\Dm G(-i\omega')\vec{G}^{(0)}_{2}\rt)_1\n\\
&\times\lt(G(-i\nu+i\omega')\Dp G(-i\nu)\vec{G}^{(0)}_{2}\rt)_1,\label{Cros-inel-pre-a}
\end{align}
\begin{align}
C^{(1;1)}(-i\nu)&=(G\vec{L})_2\lt(G(-i\nu)\Dm G\vec{L}\rt)_1\lt(G(-i\nu)\vec{G}^{(+)}_{2}(0)\rt)_1\n\\
&+\chast\lt(G(-i\nu)\Dm G(-i\omega')\vec{G}^{(0)}_{2}\rt)_1\n\\
&\times\lt(G(-i\nu+i\omega')\vec{G}^{(+)}_{2}(\omega')\rt)_1\n\\
&+\chast(G\vec{L})_1\lt(G(-i\nu)\Dm G(-i\nu)G(-i\omega')\vec{G}^{(0)}_{2}\rt)_1\n\\
&\times\lt(G(i\omega')\Dp G\vec{L}\rt)_2,\label{Cros-inel-pre-b}
\end{align}
\begin{align}
C^{(0;2)}(-i\nu)&=\chast(G\vec{L})_2\n\\
&\times\lt(G(-i\nu)G(-i\omega')\Dm G\vec{L}\rt)_1\lt(G(i\omega')\vec{G}^{(+)}_{1}(0)\rt)_2\n\\
&+\chast(G\vec{L})_1\n\\
&\times\lt(G(-i\nu)G(-i\omega')\vec{G}^{(-)}_{2}(0)\rt)_1\lt(G(i\omega')\Dp G\vec{L}\rt)_2\n\\
&+\chast\chastA\n\\
&\times\lt(G(-i\nu)G(i\omega'')\Dm G(-i\omega')\vec{G}^{(0)}_{2}\rt)_1\n\\
&\times\lt(G(-i\omega'')G(i\omega')\Dp G\vec{L}\rt)_2(G\vec{L})_1\n\\
&+\chast\chastA\lt(G(-i\omega'')\vec{G}^{(+)}_{1}(i\omega')\rt)_2\n\\
&\times\lt(G(-i\nu)G(i\omega'')\Dm G(-i\omega')\vec{G}^{(0)}_{2}\rt)_1\n\\
&+\chast\chastA\n\\
&\times\lt(G(-i\nu)G(i\omega'')\vec{G}^{(-)}_{2}(-i\omega')\rt)_1\n\\
&\times\lt(G(-i\omega'')\Dp G(i\omega')\vec{G}^{(0)}_{1}\rt)_2\n\\
&+\chast\chastA\lt(G(-i\omega'')\Dp G(i\omega')\vec{G}^{(0)}_{1}\rt)_2\n\\
&\times(G\vec{L})_2\lt(G(-i\nu)G(i\omega'')G(-i\omega')\Dm G\vec{L}\rt)_1.\label{Cros-inel-pre-c}
\end{align}

Calculation of the real part of Eqs.~(\ref{Cros-inel-pre-a},\ref{Cros-inel-pre-b},\ref{Cros-inel-pre-c}) is done by
employing the same substitutions as in the case of the background
spectrum (see above). The final result follows
after subsequent integration of 
Eqs.~(\ref{Cros-inel-pre-a},\ref{Cros-inel-pre-b},\ref{Cros-inel-pre-c}) over $\omega''$ and, partially, over
$\omega'$ by using the identities (\ref{eq:general_sum_rule}), (\ref{sum-rule-II}), and (\ref{sum-rule-III}).


\begin{thebibliography}{99}

\bibitem{sheng}
P.~Sheng, {\it Introduction to Wave Scattering, Localization and
Mesoscopic Phenomena} (Academic Press, San Diego, 1995).

\bibitem{albada85}
M.~P. Van Albada and A.~Lagendijk, Phys.~Rev.~Lett.{\bf 55}, 2692
(1985); P.-E.~Wolf and G.~Maret, Phys.~Rev.~Lett.{\bf 55}, 2696
(1985).

\bibitem{storzer06}
M.~St\"orzer, P.~Gross, C.~M.~Aegerter, and G.~Maret,
Phys.~Rev.~Lett. {\bf 96}, 063904 (2006).

\bibitem{wiersma08}
D.~S.~Wiersma, Nature Physics {\bf 4}, 359 (2008).

\bibitem{labeyrie99}
G. Labeyrie, F. de Tomasi, J.-C. Bernard, C. A. M\"uller, C. Miniatura, and R. Kaiser,  Phys.~Rev.~Lett. {\bf 83}, 5266 (1999).


\bibitem{jonckheere00}
T.~Jonckheere, C.~A.~M\"uller, R.~Kaiser, C.~Miniatura and
D.~Delande, Phys.~Rev.~Lett. {\bf 85}, 4269 (2000).

\bibitem{labeyrie03}
G.~Labeyrie, D.~Delande, C.~A.~M\"uller, C.~Miniatura and R.~Kaiser,
Europhys.~Lett. {\bf 61}, 327 (2003).

\bibitem{bidel02} Y.~Bidel, B.~Klappauf, J.~C.~Bernard, D.~Delande,
 G.~Labeyrie, C.~Miniatura, D.~Wilkowski, and R.~Kaiser,
 Phys.~Rev.~Lett. {\bf 88}, 203902 (2002).

\bibitem{scully}
M.~O.~Scully and M.~S.~Zubairy, {\it Quantum optics}, (Cambridge
University Press, Cambridge, 1997)

\bibitem{chaneliere04}
T.~Chaneli\`ere, D.~Wilkowski, Y.~Bidel, R.~Kaiser, and C.~Miniatura, Phys.~Rev.~E {\bf 70}, 036602 (2004).

\bibitem{mueller01}
C.~A.~M\"uller, T.~Jonckheere, C.~Miniatura and D.~Delande,
Phys.~Rev.~A {\bf 64}, 053804 (2001).

\bibitem{kupriyanov03}
D.~V.~Kupriyanov, I.~M.~Sokolov, P.~Kulatunga, C.~I.~Sukenik, and
M.~D.~Havey, Phys.~Rev.~A {\bf 67}, 013814 (2003).



\bibitem{balik05}
S.~Balik, P.~Kulatunga, C.~I.~Sukenik, M.~D.~Havey, D.~V.~Kupriyanov, and I.~M.~Sokolov, J.~Mod.~Opt. {\bf 52}, 2269
(2005).

\bibitem{wellens08}
T.~Wellens and B.~Gr\'emaud, Phys.~Rev.~Lett. {\bf 100}, 033902
(2008).

\bibitem{wellens09b}
T.~Wellens and B.~Gr\'emaud, Phys.~Rev.~A {\bf 80}, 063827 (2009).

\bibitem{wellens04}
T.~Wellens, B.~Gr\'emaud, D.~Delande, and C.~Miniatura, Phys. Rev. A
{\bf 70}, 023817 (2004).

\bibitem{shatokhin05}
V. Shatokhin, C.~A.~M\"uller, and A. Buchleitner, Phys.~Rev.~Lett.
{\bf 94}, 043603 (2005).

\bibitem{gremaud06}
B.~Gr\'emaud, T.~Wellens, D.~Delande, C.~Miniatura, Phys.~Rev.~A
{\bf 74}, 033808 (2006).

\bibitem{wellens06}
T.~Wellens, B.~Gr\'emaud, D.~Delande, and C.~Miniatura, Phys. Rev. A
{\bf 73}, 013802 (2006).

\bibitem{shatokhin06}
V. Shatokhin, C.~A.~M\"uller, and A. Buchleitner, Phys.~Rev.~A {\bf
73}, 063813 (2006).

\bibitem{wellens09}
T.~Wellens, T. Geiger, V. Shatokhin, and A. Buchleitner, arXiv:0909:5607.

\bibitem{geiger09}
T. Geiger, {\it New approach to multiple scattering of intense laser light from cold atoms}, Diploma thesis, Albert-Ludwigs Universit\"at Freiburg (2009),
http://www.freidok.uni-freiburg.de/volltexte/6986

\bibitem{lehmberg70}
R.~H.~Lehmberg, Phys.~Rev.~A {\bf 2}, 883 (1970).

\bibitem{agarwal74}
G.~S.~Agarwal, {\it Quantum Statistical Theories and their Relation
to Other Approaches} (Springer, Berlin, 1974).

\bibitem{shatokhin07}
V. Shatokhin, T.~Wellens, B.~Gr\'emaud, and A. Buchleitner, Phys. Rev. A {\bf 76}, 043832 (2007).

\bibitem{rist08}S. Rist, J. Eschner, M. Hennrich, and G. Morigi, Phys. Rev A {\bf 78}, 013808 (2008).

\bibitem{landau-V}
L.~D.~Landau and E.~M.~Lifshitz, {\it Statistical Physics, Course of Theoretical Physics, V} (Pergamon Press, Oxford, 1980).

\bibitem{carmichael}
H.~Carmichael, {\it An Open System Approach to Quantum Optics}, (Springer, Berlin, 1993).

\bibitem{horn}
R.~A. Horn and C.~R.~Johnson, {\it Matrix Analysis}, (Cambridge
University Press, Cambridge, 1985).

\bibitem{mollow69}
B.~R.~Mollow, Phys. Rev. {\bf 188}, 1969 (1969).

\bibitem{skornia01}C. Skornia, J. von Zanthier, G. S. Agarwal, E. Werner, and H. Walther, Phys. Rev. A {\bf 64}, 063801 (2001).


\end{thebibliography}
\end{document}